\documentclass[11pt]{article}
\usepackage{fullpage}
\usepackage{lmodern,microtype}
\usepackage{amsmath,amssymb,amsthm}
\usepackage{thmtools}
\usepackage[colorlinks=true,linkcolor=blue,citecolor=blue,urlcolor=blue]{hyperref}
\hypersetup{hypertexnames=false}

\usepackage{authblk}

\usepackage{thm-restate}
\usepackage[capitalise]{cleveref}

\theoremstyle{plain}
\newtheorem{theorem}{Theorem}[section]
\newtheorem{lemma}[theorem]{Lemma}
\newtheorem{claim}[theorem]{Claim}
\newtheorem{proposition}[theorem]{Proposition}
\newtheorem{corollary}[theorem]{Corollary}
\newtheorem{definition}[theorem]{Definition}
\newtheorem{example}[theorem]{Example}
\newtheorem{remark}[theorem]{Remark}
\usepackage{physics}
\newcommand{\E}{\mathop{\mathbb{E}}\displaylimits}
\newcommand{\R}{\mathbb R}
\newcommand{\Z}{\mathbb Z}
\newcommand{\cD}{\mathcal D}
\newcommand{\falling}[2]{(#1)_{#2}}
\newcommand{\eps}{\varepsilon}
\DeclareMathOperator{\dist}{dist}

\usepackage{xcolor}
\usepackage[disable,colorinlistoftodos,textsize=tiny]{todonotes}

\newcommand{\nocontentsline}[3]{}
\newcommand{\tocless}[2]{\bgroup\let\addcontentsline=\nocontentsline#1{#2}\egroup}

\usepackage{array}
\usepackage{booktabs}

\begin{document}
\title{Quantum Query Lower Bounds from Classical Moment Matching}

\author{Chandrima Kayal\thanks{Email: \texttt{chandrimakayal2012@gmail.com}}}
\author{Fr\'ed\'eric Magniez\thanks{Email: \texttt{frederic.magniez@irif.fr}}} 
\affil{CNRS, Université Paris Cité, IRIF, France}
\author{Sayantan Sen\thanks{Email: \texttt{sayantan789@gmail.com}}}
\affil{Centre for Quantum Technologies, National University of Singapore, Singapore}
\author{D\'aniel Szab\'o \thanks{Email: \texttt{szabo.daniel@lmu.de}}}
\affil{Ludwig-Maximilians-Universität München \& MCQST, Munich, Germany}

\date{}
    \pagenumbering{roman}
	\maketitle

\thispagestyle{empty}

\begin{abstract}
We prove a quantum analogue of the classical moment matching method, a popular framework used to prove lower bounds in property testing. For two random variables $X_0$ and $X_1$ whose moments match up to the first $r-1$ moments, the classical moment matching method says that no classical algorithm can reliably distinguish between input instances of length $N$ constructed from $X_0$ and $X_1$, respectively, unless it performs at least $\Omega(N^{1-1/r})$ queries. While this method has been widely explored since its inception in the seminal work of Raskhodnikova, Ron, Shpilka, and Smith (STOC 2007) to prove lower bounds for symmetric properties, it has remained an open question whether a quantum analogue of it could be proved.

We prove two variants of this quantum moment matching method, which are incomparable. Our first quantum query lower bound scales as $\Omega(N^{1/2-1/(2r)})$, while our second moment matching theorem yields a sharper lower bound of $\Omega(N^{1/2-1/(2(2^r-1))})$, albeit for smaller parameter ranges. Our results combine a new continuous hybrid argument with the polynomial method. The continuous hybrid argument compares two hard input distributions generated from i.i.d. samples by continuously varying both distributions, keeping them identically distributed at every intermediate stage. It reduces the comparison to the effect of changing the distribution of a single sample. The polynomial method then cancels the terms determined by the matched moments.

We apply these moment matching theorems to prove new optimal lower bounds on $k$-distinctness, $k$-star-freeness, planarity and outerplanarity testing, as well as to improve prior lower bounds on image size testing and $K$-junta testing by logarithmic factors. A key advantage of our framework is its simplicity: it reduces quantum lower bound proofs to the construction of suitable moment-matching distributions, which in many cases have already been constructed for establishing corresponding classical lower bounds.
\end{abstract}

\thispagestyle{empty}
\clearpage

\renewcommand{\baselinestretch}{0.95}\normalsize
\tableofcontents
\thispagestyle{empty}
\clearpage
\renewcommand{\baselinestretch}{1.0}\normalsize

\pagenumbering{arabic}

\section{Introduction}\label{sec:introduction}

\subsection{Context and motivations}

\paragraph{Property testing}
In this era of big data, even just reading the input in its entirety can be prohibitively costly. However, from a practical perspective, we can often tolerate some sort of approximation. The field of \emph{property testing} focuses on such a setting: here the goal is to distinguish whether the input object (in this work a function or a graph) satisfies some property of interest, or is \emph{far} from all such objects under some suitable distance measure, by accessing the input as few times as possible. The tester can either accept or reject when the distance of the input from the property does not satisfy the condition.  Perhaps surprisingly, it is often possible to design testers whose complexity scales sublinearly with the size of the input. Over the last two and a half decades, property testing has become a vast and active area of research, see \cite{goldreich2017introduction, bhattacharyya2022property,DBLP:journals/fttcs/Ron09,DBLP:conf/propertytesting/CzumajS10,DBLP:journals/siamdm/RubinfeldS11,canonne2020survey, canonne2022topics} for more details.

Property testing has naturally also been explored in the quantum setting~\cite{friedl2003quantum,buhrman2003quantum}. For several important problems such as junta testing, graph isomorphism, bipartiteness and expansion testing, polynomial quantum speedups are known~\cite{ambainis2011quantum,ABRW16,AMSS25,DBLP:conf/tqc/ChenAH24}, see the survey~\cite{montanaro2016survey} for a detailed discussion. Naturally, understanding the limit of quantum advantage has been a prime focus in the community.

Our starting point is a function $f:[N]\to[N]$, where
$[N]=\{1,\ldots,N\}$. For a property (aka. collection) $\mathcal P$ of such functions,
the input is \emph{$\varepsilon$-far} from $\mathcal P$ if at least
$\varepsilon N$ values must be changed to obtain a function in
$\mathcal P$. Given oracle access to an unknown function $f$, an \emph{$\varepsilon$-tester} must distinguish whether $f \in \mathcal{P}$, or $\varepsilon$-far from $\mathcal{P}$ with success probability at least $2/3$, by performing as few queries as possible to the oracle of $f$. 
A classical query reveals $f(i)$ at a chosen index $i\in[N]$, while a quantum algorithm has coherent access to $f$, and can query several indices in quantum superposition.

Many natural properties of interest, such as injectivity, the absence of \(k\)-collisions, or having an image of at least a given size, depend only on how many times each output value occurs, i.e., on their frequencies (aka. multiplicities). Writing
\[
 c_y=|\{i\in[N]:f(i)=y\}|,
\]
the image size of \(f\) is the number of \(y\) with \(c_y>0\), and \(f\) has a \(k\)-collision precisely when some \(c_y\ge k\). The vector $\mathbf c=(c_1,c_2,\ldots)$ is called the \emph{frequency vector} of $f$.
Indeed, these properties remain invariant under permutations of the input indices and the output
labels. Such properties are called \emph{symmetric} or \emph{label-invariant} properties and play an important role in both classical and quantum property testing, and have been extensively studied in the literature, see \cite{valiant2008testing,valiant2011power,valiant2017estimating,wu2019chebyshev, DBLP:journals/tit/ChakrabortyFGMS25} and the references therein.

Since these symmetric properties only depend on the multiplicities of the output labels, a natural approach to prove lower bounds for such properties is to consider hard instances based on frequencies and then randomize the positions and labels. Thus the task is to construct two hard input instances (or distributions of instances), one in the property $\mathcal{P}$, and another far from $\mathcal{P}$, and prove that any algorithm performing only a few queries cannot reliably distinguish them.

\paragraph{Moment-matching method}
Moment-matching method is a powerful classical framework that is used to achieve this dichotomy. 
In general, for some symmetric property $\mathcal{P}$, a pair of input instances with one instance in the property $\mathcal{P}$ and the other being far from $\mathcal{P}$, and whose moments match up to $(r-1)$ moments for some integer $r$, would require $\Omega(N^{1-1/r})$ queries to distinguish them with probability at least $2/3$.
To build its intuition, let us consider the example of distinguishing a uniformly random $1$-to-$1$ function $f_0: [N] \rightarrow [N]$ from a uniformly random $2$-to-$1$ function $f_1: [N] \rightarrow [N]$. Let us choose a label $Y$ uniformly from $[N]$, and let $Z_b=|f_b^{-1}(Y)|$ for $b \in \{0,1\}$. Note that $Z_0=1$ with probability $1$ and $Z_1$ is either $0$ or $2$, each with probability $1/2$.
Then the moment-matching method proves that distinguishing between these two cases with probability at least $2/3$ requires $\Omega(N^{1-1/2})$ queries. 

The intuition for this framework stems from the fact that since $\mathcal{P}$ is symmetric, \emph{moments are all that matter}.
One can alternatively define $X_0$ and $X_1$ for the non-zero frequencies. In that case, $Y$ is uniformly sampled from $f([N])$. Then $Y_0=1$ and $Y_1=2$ with probability $1$ for our example.
One can show that exact moment matching for $Z_0,Z_1$ can be translated to proportional moment matching for $X_0,X_1$, and this is usually the expression that is used in the literature.

The classical moment matching framework was introduced in the seminal work of \cite{RRSS09}, where the authors used this framework to prove lower bounds on estimating the support size and the entropy of a distribution, and the number of distinct elements in a sequence. Then \cite{valiant2011power} extended this framework to approximately matching moments and proved lower bounds on various distribution testing problems such as distance estimation. Later, Wu and Yang~\cite{wu2019chebyshev} connected the
matching moments framework to polynomial
approximation and obtained sharp bounds for
entropy and support-size estimation. Since then, there have been a flurry of results in the context of proving distribution testing lower bounds that employ this framework~\cite{AOST17,canonne2022price,diakonikolas2018sharp,goldreich2023lower,canonne2025settling}. The same framework was used to prove a $k$-collision-freeness lower bound
by Hellweg and Sohler~\cite{HS12} for $k=3$, which was later extended to arbitrary $k$ by Peng and
Wang~\cite{peng2023optimal}, which they transfer to directed subgraph-freeness via a reduction.

Notably, there are two important ingredients of this framework: (i) when using it, one has to construct suitable random variables so that the framework can be applied; and (ii) the proof of the framework provides that no algorithm can reliably distinguish them by performing only a few queries. While these series of works use different input instance constructions and moment conditions, all of these proofs use the \emph{Poissonization} trick: instead of considering a fixed number of samples, taking a set of samples distributed according to the Poisson distribution makes the frequencies of different symbols independent, thereby greatly simplifying the proofs.

\paragraph{Toward a quantum analogue}
Due to the wide success of this framework and its simplicity to use in practice for proving new lower bounds, it was naturally considered whether a similar lower bound framework exists for 
the quantum query model.  Ambainis, Belovs, Regev, and de~Wolf~\cite{ABRW16} were the first to propose whether combining the construction of \cite{RRSS09} with quantum lower-bound techniques could be used to obtain stronger bounds for junta testing~\cite{ABRW16}. %
Without answering directly to this question, several problems, for which the moment matching frameworks can be applied classically, were studied in the quantum setting using mostly the dual polynomial method and quite involved mathematical constructions.
First, improved lower bounds for junta testing and several other important problems, like image size testing, were achieved in a subsequent work~\cite{BKT20}. Later on, \cite{AMSS25} establishes lower bounds for collision-freeness and directed subgraph-freeness testing, which was later extended by \cite{PW26}. 
Thus, quantum lower bounds of some specific problems were tackled, but not directly the general matching moments method.
We now briefly discuss the bottleneck.

Recall that the second ingredient of matching-moments-based lower bound proofs is to prove \emph{indistinguishability}: a classical algorithm that performs too few queries cannot reliably distinguish between the two hard distributions.  Since the use of Poissonization makes the number of occurrences of every label independent, it allows to compare the histogram distributions based on the input distributions. Since the property is symmetric, the algorithm's decision depends only on this
histogram. Consequently, if two histogram distributions are close in total variation distance, no algorithm making only a few queries can distinguish them with high probability.%

However, a quantum algorithm can query several indices of the input in superposition, without measuring the answer register after each query. In fact, measuring the outcome of each query can remove coherence used by subsequent queries and thereby change the algorithm's acceptance probability. Thus, even for symmetric properties, a quantum query algorithm cannot be analyzed by merely processing a classical histogram constructed from its samples. A quantum moment-matching argument must therefore control what coherent queries can distinguish, which has not been studied in the literature. Thus, despite a long series of works, the following question remained wide open: %
\begin{center}
    \emph{Does there exist a quantum moment matching framework for proving quantum query lower bounds?}
\end{center}

\subsection{Our results}\label{sec:ourresults}

In this work, we answer the above question in the affirmative:
\begin{center}
    \emph{There exists a quantum moment matching framework for proving quantum query lower bounds.}
\end{center}
We prove that indeed such a general lower bound framework exists in the quantum query realm,
transferring each corresponding classical bound $\Omega(N^{1-1/r})$ to the quantum bound
$O(N^{1/2-1/(2r)})$, and even sometimes to the higher bound $\Omega(N^{1/2-1/(2(2^r-1))})$.

\paragraph{Quantum analogue of classical moment matching framework} 
We provide two variants of the moment matching theorem: the first one works for a large range of parameters, useful for broader applicability, whereas the second one, while restricted to smaller parameter ranges, yields better bounds on several problems of interest. Let us start with the first variant of our quantum moment matching framework.

In both theorems, we start as in the classical setting.
Fix an integer $A\ge1$. Suppose $X_0,X_1$ are two random variables supported on $[A]$ such that for an integer $r\ge2$, 
\begin{equation} \label{eq:proportionalintro}
 \frac{\E\,[X_1]}{\E\,[X_0]}
 =\frac{\E\,[X_1^2]}{\E\,[X_0^2]}
 =\cdots
 =\frac{\E\,[X_1^{r-1}]}{\E\,[X_0^{r-1}]}.
\end{equation}
These random variables refer to the statistics of the frequency vector of any candidate function in the hard distributions. Indeed from them, and any given integer $N\geq 1$, one can construct a pair of 
hard distributions $\Pi_{0,N}$ and $\Pi_{1,N}$ (constructed from $X_0$ and $X_1$, respectively) of functions $f:[N]\to[N]$.
\begin{theorem}[Informal statement of \cref{thm:proportional-polynomial}]\label{thm:proportional-polynomialintro}
Assume that $X_0,X_1$ satisfy Equation~\ref{eq:proportionalintro}.
When $A=O(r^2)$ and $A=o(\sqrt{N/\log N})$,
any $T$-query quantum algorithm can only distinguish between $\Pi_{0,N}$ and $\Pi_{1,N}$ with probability at least $2/3$ whenever $T=\Omega(N^{1/2-1/(2r)})$.
\end{theorem}

Note that the above result gives a quadratic decrease in the lower bound compared to the classical setting, reflecting the possibility of quantum speedups.
The above theorem is similar to the matching moment theorem of \cite{RRSS09}, and it holds for a large range of parameters $A$ and $r$ (see \Cref{thm:proportional-polynomial} for the exact ranges of the parameters). We employ this theorem to prove optimal quantum query lower bounds on 
image size testing and junta testing.  

However, for several properties like $k$-distinctness and subgraph-freeness testing, the bound obtained from the above theorem remains sub-optimal. To achieve the optimal lower bounds for these properties, we further improve the above theorem to give tighter bounds, albeit for more restricted parameter ranges. %
\begin{theorem}[Informal statement of \Cref{thm:proportional-differences}]\label{thm:proportional-differencesintro}
Assume that $X_0,X_1$ satisfy Equation~\ref{eq:proportionalintro}.
For fixed $A,r$,
any $T$-query quantum algorithm can only distinguish between $\Pi_{0,N}$ and $\Pi_{1,N}$ with probability at least $2/3$ whenever $T=\Omega(N^{1/2-1/(2(2^r-1))})$.
\end{theorem}

A crucial appeal of this framework lies in its simplicity for proving new quantum query lower bounds. These theorems can be used as \emph{black-box}: all it needs is input instances that satisfy the conditions of the theorems, and we directly obtain a new quantum query lower bound! This is indeed the case of various applications. In fact, in many cases, it is sufficient to consider the classical moment-matching lower bound construction, and even designing new hard instances is not required.

\paragraph{Applications of the quantum moment matching framework}

As applications of our two moment matching theorems, we obtain several new quantum query lower bounds, even optimal ones, with either polynomial or logarithmic improvements. Moreover, all these lower bounds are now proven in a more direct and streamlined fashion.
Our results are summarized in \Cref{tab:applications-summary}. 

\begin{table}[!htb]
\centering
\small
\setlength{\tabcolsep}{3pt}
\renewcommand{\arraystretch}{1.15}

\begin{tabular}{@{}
 >{\raggedright\arraybackslash}p{\dimexpr.24\linewidth-1.5\tabcolsep\relax}
 >{\raggedright\arraybackslash}p{\dimexpr.25\linewidth-1.5\tabcolsep\relax}
 >{\raggedright\arraybackslash}p{\dimexpr.28\linewidth-1.5\tabcolsep\relax}
 >{\raggedright\arraybackslash}p{\dimexpr.23\linewidth-1.5\tabcolsep\relax}
 @{}}
\toprule
Property testing problem & \multicolumn{2}{c}{Previous bounds} & \multicolumn{1}{c}{This work}  \\
\cmidrule(lr){2-3}\cmidrule(l){4-4}
& Classical & Quantum & Quantum \\
\midrule

$k$-distinctness
& $\Omega(N^{1-1/k})$\par\cite{peng2023optimal}
& $\Omega\!\left(\dfrac{N^{1/2-1/(2k)}}{\log^2N}\right)$
  \par\cite{AMSS25}
& $\Omega(N^{a_k})$\par
  Corollary~\ref{cor:k-distinctness} \\
& $O(N^{1-1/k})$\par\cite{HS12}
& $O(N^{a_k})$\par\cite{AMSS25}
& \\
\addlinespace

$k$-star-freeness and $k$\nobreakdash-source-subgraph-freeness %
& $\Omega(N^{1-1/k})$\par\cite{peng2023optimal}
& $\Omega\!\left(\dfrac{N^{1/2-1/(2k)}}{\log^2N}\right)$
  \par\cite{AMSS25}
& $\Omega(N^{a_k})$\par
  Corollary~\ref{s2:thm:directedgraph} \\
& $O(N^{1-1/k})$\par\cite{HS12}
& $O(N^{a_k})$\par\cite{AMSS25}
& \\
\addlinespace

Planarity and outerplanarity
& $\Omega(N^{2/3})$\par\cite{peng2023optimal}
& $\Omega(N^{1/3})$\par
  consequence of~\cite{Zhandry15}
& $\Omega(N^{3/7})$\par
  Corollary~\ref{s2:thm:directedgraph} \\
& $O(N)$
& $O(N)$
& \\
\addlinespace

Image-size testing
& $\Omega(N^{1-\delta})$\par
  any fixed $\delta>0$\par\cite{RRSS09}
& $\Omega\!\left(\dfrac{\sqrt N}{\log^2N}\right)$
  \par\cite{BKT20}
& $\Omega(\sqrt N)$\par
  Corollary~\ref{cor:image-size} \\
& $O(N/\log N)$\par\cite{wu2019chebyshev}
& $O(\sqrt N)$\par\cite{montanaro2015quantum}
& \\
\addlinespace

$K$-junta testing
& $\Omega(K\log K)$\par\cite{sauglam2018near}
& $\Omega\!\left(\dfrac{\sqrt K}{\log^2K}\right)$
  \par\cite{BKT20}
& $\Omega(\sqrt K)$\par
  Corollary~\ref{s3:thm:junta} \\
& $O(K\log K)$\par\cite{blais2009testing}
& $O(\sqrt K\,\log K)$\par\cite{ABRW16}
& \\

\bottomrule
\end{tabular}

\caption{Query bounds for our applications, with explicit logarithmic
factors, $a_k=(2^{k-1}-1)/(2^k-1)$. For each problem, the previous bounds occupy two rows:
lower bounds above, upper bounds below. This work gives the lower
bounds in the last column.
The $k$-distinctness and directed subgraph-freeness entries are
for $k\ge3$.
In the image-size testing row, $\delta>0$ is an arbitrary constant. 
}
\label{tab:applications-summary}
\end{table}

While \Cref{thm:proportional-differencesintro} proves a sharper lower bound compared to  \Cref{thm:proportional-polynomialintro}, as we mentioned, \Cref{thm:proportional-differencesintro} holds for smaller parameter ranges. For our applications of image size testing and junta testing, applying \Cref{thm:proportional-differencesintro} would not yield optimal lower bounds and \Cref{thm:proportional-polynomialintro} needs to be applied. On the other hand, for $k$-distinctness, directed $k$-star-freeness and directed planarity and outerplanarity testing, \Cref{thm:proportional-differencesintro} is required for proving the optimal bounds. Thus, \Cref{thm:proportional-polynomialintro} and \Cref{thm:proportional-differencesintro} are incomparable from an application standpoint.

We start with the \emph{$k$-distinctness} testing problem. Here, given access to a function $f:[N]\to[M]$, the goal is to distinguish whether every output value of $f$ occurs fewer than $k$ times or one needs to change the value of at least $\varepsilon N$ positions to make it so. This is also often called \emph{$k$-collision-freeness} testing. 
Next we consider four problems in directed bounded-degree graphs in the \emph{unidirectional model}, where the algorithm can only query the outgoing neighbors of each vertex. The first two problems are testing \emph{$k$-star-freeness} and \emph{$k$-source-subgraph-freeness}: whether a $k$-star or any given subgraph with $k$ source components appear as subgraphs in the graph (see \Cref{s2:sec:graph-model} for the definition).
Then, we consider planarity and outerplanarity testing: given query access to a graph in the unidirectional model, the algorithm has to distinguish whether the underlying undirected graph is \emph{planar} or \emph{outerplanar}, or far from being so. Our results for these problems are summarized in the following theorem, see \Cref{tab:applications-summary} for the best prior classical and quantum bounds.
\begin{theorem}[Informal statement of Corollary~\ref{cor:k-distinctness} and  Corollary~\ref{s2:thm:directedgraph}]
The quantum query complexity of property testing: 
\begin{itemize}
    \item[(i)] $k$-distinctness for strings of length $N$ is $\Omega(N^{(2^{k-1}-1)/(2^k-1)})$;

    \item[(ii)] $k$-star-freeness and $k$-source-subgraph-freeness for $N$-vertex graphs in the unidirectional model is $\Omega(N^{(2^{k-1}-1)/(2^k-1)})$;

    \item[(iii)] planarity and outerplanarity of $N$-vertex graphs in the unidirectional model is $\Omega(N^{3/7})$.
\end{itemize}
\end{theorem}

Suppose we are given query access to a function $f:[N]\to[M]$.
In the \emph{image size testing} problem, our goal is to distinguish whether $f$ takes at most a prescribed number of distinct values, or is far from all such functions. Similarly, in $K$-junta testing, our goal is to test if $f$ depends on at most $K$ input coordinates among its $N$ coordinates. Such a function admits a low-dimensional representation and has been widely studied in the literature. Our results are stated as follows:
\begin{theorem}[Informal statements of Corollaries~\ref{cor:image-size} and \ref{s3:thm:junta}]
The quantum query complexity of
\begin{itemize}
    \item[(i)] Image size testing for strings of length $N$ is $\Omega(\sqrt{N})$,

    \item[(ii)] $K$-junta testing is $\Omega(\sqrt{K})$.
\end{itemize}
\end{theorem}

The best known prior classical and quantum bounds on these problems are presented in \Cref{tab:applications-summary}. 
A detailed discussion of techniques and prior works on these applications is presented in \Cref{sec:applications}.

\subsection{Technical overview}
We first give an overview of the proof of \Cref{thm:proportional-polynomialintro},
then move to \Cref{thm:proportional-differencesintro}.
Our proofs combine the polynomial method with a continuous hybrid
argument. We express acceptance probabilities as polynomials and
use the hybrid argument to compare their expectations under the
hard input distributions.

Polynomial and discrete hybrid arguments have previously been
combined, for instance in Zhandry's construction of quantum-secure
pseudorandom functions~\cite{Zhandry12}, but not so often applied since.
Here, continuous interpolation preserves the i.i.d.\ structure of
our multiple samples, while moment matching cancels the
low-degree contributions at each stage.

\subsubsection{First step: input distributions}\label{sec:introinputdistributions} %
We start with two random variables $X_0,X_1$ supported on $[A]=\{1, \ldots, A\}$ that correspond to \emph{positive frequency statistics} and have proportional moments up to order $r-1$. Let $\mu_b=\E\,[X_b]$ for $b \in \{0,1\}$. 
From $X_b$ we define a related {mean-one frequency statistics} $Z_b$ as follows: keep a sample from $X_b$ with probability $1/\mu_b$ and replace it by zero otherwise. The resulting random variable $Z_b$ includes zero frequencies and has support on $S=\{0\}\cup[A]$.
It also satisfies
\[
 \E\,[Z_b]=1,\qquad
 \E\,[Z_b^j]=\frac{\E\,[X_b^j]}{\mu_b},
 \qquad
 \E\,[Z_0^j]=\E\,[Z_1^j]\quad(1\le j<r).
\]
The additional zero value is crucial, it lets us obtain mean one while converting
\emph{proportional} moments into \emph{matching} moments.

For $b \in \{0,1\}$, we now construct a random function from prescribed positive frequency statistics $X_b$, or equivalently, its associated mean-one frequency statistics $Z_b$. 
Let $\pi_b$ be the underlying distribution of $Z_b$. 
This is not as easy as defining statistics from a given function $f$. %
The main difficulty is that we would like a sampling process that samples $f$ with as many i.i.d samples from $\pi_b$ as possible. However, for an input $f:[N]\to[N]$, the sum of the frequencies has to be exactly $N$, and because of the randomness of the samples we cannot be sure to achieve this after any number of samples from $\pi_b$.

To overcome this issue, we will use a randomized variant of the frequency construction of~\cite{RRSS09} %
to define $\Pi_{b,N}$ from $m=\lfloor N/2\rfloor$ i.i.d. samples of $\pi_b$. 
\begin{enumerate}
\item We sample $m=\lfloor N/2\rfloor$ independent multiplicities
      $\mathbf z=(z_1,\ldots,z_m)$ according to the distribution $\pi_b$ of $Z_b$; each nonzero draw specifying the multiplicity of a distinct output label. Set $z=\sum_i z_i$. 
\item If $z\le N$, we add $N-z$ new output values that occur once each (aka. \emph{singletons}) yielding multiplicity vector $\mathbf{c}$, and assign output labels and input positions uniformly at random.
\item If $z>N$, $\mathbf{c}$ is set to be the all-$1$ vector, thus after taking uniformly random labels we obtain a uniformly random bijection.
\end{enumerate}
This gives a function on exactly $N$ inputs. The construction ensures that the original
multiplicity samples are independent. For this construction, we say that we draw $f\sim\cD_{\mathbf c}$ where \(\mathcal D_{\mathbf c}\) is the uniform distribution over functions with the prescribed output multiplicities $\mathbf{c}$, up to relabeling. This complete randomized procedure defines the distribution $\Pi_{b,N}$ over functions. After sampling \(f\sim\Pi_{b,N}\), we provide the quantum algorithm \(\mathcal A\) with oracle access to that fixed function. Therefore, its average acceptance probability is  $\Pr_{f\sim\Pi_{b,N}}[\mathcal A^f=1]$.

It is important to note that our construction differs slightly from the original one of~\cite{RRSS09}.  Indeed, \cite{RRSS09} obtains independence of the
frequency statistics of the observed values, rather than of the input multiplicities. %
For this, they use a Poissonization argument: instead of taking $T$ samples to answer $T$ queries, they consider a Poisson($T$) number of such samples.%

\subsubsection{Second step: forming the acceptance probability polynomial}
\label{sec:first-moment-overview}
Our main goal is to show that any quantum algorithm that performs few queries has
almost the same acceptance probability under the two input distributions whose moments match up to $r-1$ moments. We will use the \emph{polynomial method}, i.e., the fact that the acceptance probability of a $T$-query quantum algorithm is a polynomial of degree at most $2T$
in the indicators of the oracle values~\cite{BBCMW01}. 

Averaging over the random assignments of positions and
labels gives a symmetric polynomial $P$ of degree at most $2T$ in
the label multiplicities. More precisely,
Lemma~\ref{lem:symmetrization} gives
\begin{equation}\label{eq:introP}
 P(\mathbf c)
 =\Pr_{f\sim\cD_{\mathbf c}}[\mathcal A^f=1]\in[0,1]
\end{equation}
for every feasible frequency vector $\mathbf c$.
Writing $\mathbf c(\mathbf z)$ for the frequency vector produced
by our construction from the sampled multiplicities $\mathbf z$,
including the fallback case, we obtain
\[
 \Pr_{f\sim\Pi_{b,N}}[\mathcal A^f=1]
 =\E_{\mathbf z\sim\pi_b^{\otimes m}}[P(\mathbf c(\mathbf z))].
\]
Symmetry means that permuting the entries of $\mathbf c$ leaves
$P(\mathbf c)$ unchanged. It also allows us to express $P$ in
terms of the power sums $p_j(\mathbf c)=\sum_{a=1}^N c_a^j$,
connecting the acceptance probability to the frequency moments.
However, the critical point is that it is not yet necessarily a polynomial in $\mathbf{z}$.

However, we assume that $Z_0$ and $Z_1$ match only their first $r-1$ moments, while $P$ may have degree as large as $2T$, with $T$ much larger than $r$. Thus, moment matching does not directly bound the difference between the two acceptance probabilities.

The main idea is therefore to use a hybrid argument, changing the distribution of one sampled multiplicity at a time, and approximate the acceptance probability, as a function of that multiplicity with the others fixed, by a polynomial of degree below $r$. The expectations of this approximation under $\pi_0$ and $\pi_1$ agree exactly by moment-matching, and we show that the remaining error is small.

\subsubsection{Third step: using the hybrid argument}
We now describe the  \emph{hybrid argument} to analyze the success probability of any given algorithm against two different input distributions.
Recall that $\mathbf c(\mathbf z)$ is the frequency vector obtained
from the sampled multiplicities $\mathbf z$ by our input construction,
including singleton completion and the fallback when $\sum_i z_i>N$.
For $S=\{0\}\cup[A]$, define $h:S^m\to[0,1]$ by
\[
 h(\mathbf z)=P(\mathbf c(\mathbf z))
 =\Pr_{f\sim\cD_{\mathbf c(\mathbf z)}}[\mathcal A^f=1].
\]
Thus $h(\mathbf z)$ is the average acceptance probability for
fixed sampled multiplicities $\mathbf z$. 
Since permuting $\mathbf z$ only permutes the entries of $\mathbf c(\mathbf z)$, the symmetry of $P$ implies that $h$ is invariant under permutations of its coordinates. Moreover,
\[
 \Pr_{f\sim\Pi_{b,N}}[\mathcal A^f=1]
 =\E_{\mathbf z\sim\pi_b^{\otimes m}}[h(\mathbf z)].
\]

For two probability distributions $\pi_0,\pi_1$ on $S$, define the usual hybrids
\[
 H_i=\E_{\mathbf z\sim \pi_1^{\otimes i}\otimes\pi_0^{\otimes(m-i)}}[h(\mathbf z)]
 \quad(0\le i\le m).
\]
Define $q_{\mathbf w}(s)=h(\mathbf w,s)$, for $\mathbf w\in S^{m-1}$ and $s\in S$.
We are interested in the quantity $H_m-H_0$, i.e., the difference of the expected acceptance probabilities when the whole input is sampled according to $\pi_0$ or $\pi_1$.

The usual hybrid changes one coordinate at each step and telescopes
$H_m-H_0=\sum_{i=0}^{m-1}(H_{i+1}-H_i)$.
By symmetry, each increment is
\[
 H_{i+1}-H_i
 =\E_{\mathbf w\sim\pi_1^{\otimes i}\otimes\pi_0^{\otimes(m-1-i)}}
 \left[\sum_{s\in S}(\pi_1(s)-\pi_0(s))q_{\mathbf w}(s)\right].
\]

For the formal proof we use a continuous version of this argument in Lemma~\ref{lem:product-comparison}, where $(\pi_u)_{0\le u\le1}$ is a continuous, piecewise continuously differentiable path from $\pi_0$ to $\pi_1$. Setting $F(u)=\E_{\mathbf{z}\sim \pi_u^{\otimes m}}[h(\mathbf{z})]$ and  $\dot\pi_u(s)=\frac{d}{du}\pi_u(s)$, we get
\begin{equation}\label{eq:hybridintro}
|H_m-H_0| = |F(1)-F(0)| 
 \le\int_0^1|F'(u)|\,du 
 = m\int_0^1
 \E_{\mathbf w\sim\pi_u^{\otimes(m-1)}}
 \left[ \abs{\sum_{s\in S}\dot\pi_u(s) q_{\mathbf w}(s)}\right]\,du. 
\end{equation}
Then the RHS is upper bounded by $ \sum_{i=0}^{m-1}\abs{H_{i+1}-H_i}$, for the linear path
$\pi_u=(1-u)\pi_0+u\pi_1$. 

The advantage of the continuous formulation is that the remaining multiplicities $\mathbf w$ are i.i.d.\ with common distribution $\pi_u$ at every intermediate step. For each $s\in S$, the number of coordinates of $\mathbf w$ equal to $s$ therefore follows $\operatorname{Bin}(m-1,\pi_u(s))$, so standard binomial estimates apply directly. In the discrete increment $H_{i+1}-H_i$, these multiplicities remain independent but have two different distributions, $\pi_1$ and $\pi_0$. The corresponding number is then a sum of two binomial variables. Preserving the i.i.d.\ structure thus simplifies the averaging of estimates involving $q_{\mathbf w}$.

\subsubsection{Fourth step: using the matching moments}
\paragraph{Bringing back the degree bound}
In the rest of this overview, we only consider samples $\mathbf w\in S^{m-1}$ such that
$w=\sum_{i=1}^{m-1}w_i\le\frac{3N}{4}$. The opposite event has exponentially smally probability to happen and can be bounded separately using  Hoeffding’s inequality.

For every integer $0\le s\le H=\lfloor N/4\rfloor$, we have $w+s\le N$. Hence the completion procedure discussed in \Cref{sec:introinputdistributions} does not overflow and the resulting frequency vector $\mathbf c(\mathbf w,s)$ contains one $s$-tuple and $(N-w-s)$ singletons added to $\mathbf{w}$. Therefore, each $j$-th power sum of it satisfies
$p_j(\mathbf c(\mathbf w,s))=p_j(\mathbf w)+(N-w-s)+s^j$. 

Since we start with a polynomial $P$ in $\mathbf{c}$ in \Cref{eq:introP},
the function $q_{\mathbf w}(s)$ is also a polynomial in $s$ with the same degree bound as $P$.
It takes values in $[0,1]$ for every integer $0\leq s\le H$.

\paragraph{Matching moments}
Since $q_{\mathbf w}(s)$ is a degree-$2T$ polynomial in $s$, 
let $Q_{\mathbf w}(s)$ be its Taylor-truncation up to degree $r-1$.
The moment-matching assumption can be used  to cancel the contribution of $Q_{\mathbf w}$
in \Cref{eq:hybridintro}.
Indeed, since $\pi_0$ and $\pi_1$ have matching moments below order $r$, so do have 
$\pi_u$ and $\pi_v$, for any $u,v\in[0,1]$. As a consequence
$\sum_{s\in S}\pi_u(s)Q_{\mathbf w}(s)$ in fact does not depend on $u$ (specific case of \Cref{lem:moment-cancellation}).
Therefore \(
\sum_{s\in S}\dot\pi_u(s)Q_{\mathbf w}(s)=0
\) and
$$ \sum_{s\in S}\dot\pi_u(s)q_{\mathbf w}(s)
 =\sum_{s\in S}\dot\pi_u(s)
                  \bigl(q_{\mathbf w}(s)-Q_{\mathbf w}(s)\bigr).
$$

As a consequence we get this special case of \Cref{lem:path-moments}:
\[
|F(1)-F(0)| 
\le  m\int_0^1
 \E_{\mathbf w\sim\pi_u^{\otimes(m-1)}}
 \left[ \abs{\sum_{s\in S}\dot\pi_u(s) \big(q_{\mathbf w}(s)-Q_{\mathbf w}(s)\big)}\right]\,du. 
\]

\paragraph{Combining small degree and matching moments}

In the regime $N\ge\max\{8,4A\}$ and $T^2\le N/64$, Taylor approximation (Lemma~\ref{lem:taylor})  provides the approximation of the tail of $\abs{q_{\mathbf{w}}(s)-Q_{\mathbf{w}}(s)}$, using that $q_{\mathbf{w}}(s)\in[0,1]$ for every integer $0\leq s\leq H$. After taking its expectation over $\pi_b$, this leads to 
\[
\mathbb E_{Z_b\sim\pi_b}
\left[|q_{\mathbf w}(Z_b)-Q_{\mathbf w}(Z_b)|\right]
\le
\frac{2A^{r-1}}{(2r)!}
\left(\frac{128T^2}{N}\right)^r.
\]

Therefore
\[
 \begin{aligned}
 \left|\sum_{s\in S}\dot\pi_u(s)
             \bigl(q_{\mathbf w}(s)-Q_{\mathbf w}(s)\bigr)\right|
 &\le\sum_{s\in S}(\pi_0(s)+\pi_1(s))
       |q_{\mathbf w}(s)-Q_{\mathbf w}(s)|\\
 &=\sum_{b=0}^1\sum_{s\in S}\Pr[Z_b=s]\,
       |q_{\mathbf w}(s)-Q_{\mathbf w}(s)|\\
 &=\sum_{b=0}^1
       \E[|q_{\mathbf w}(Z_b)-Q_{\mathbf w}(Z_b)|]\\
 &\le\frac{4A^{r-1}}{(2r)!}
       \left(\frac{128T^2}{N}\right)^r.
 \end{aligned}
\] %

So finally
$$\abs{\Pr_{f\sim\Pi_{1,N}}[\mathcal A^f=1]-\Pr_{f\sim\Pi_{0,N}}[\mathcal A^f=1]}
=\abs{F(1)-F(0)} \leq 
\frac N2 \left[ \frac{4A^{r-1}}{(2r)!} \left(\frac{128T^2}{N}\right)^r
\right],
$$
up to an additional exponential small term (bounded by $2e^{-N/(4A^2)}$) due to our original assumption that $w\leq 3N/4$.
Thus, when $A=O(r^2)$ and $A=o(\sqrt{N/\log N})$,
this quantity can only be large when the number of queries $T=\Omega(N^{1/2-1/(2r)})$.

\subsubsection{A better bound for some regimes} %
For proving the second theorem, we use two main lemmas: Lemma \ref{ter:lem:regularization} and \Cref{ter:lem:average}. First we discuss how to prove the theorem using these lemmas and then continue with the lemmas. 

As before, let $\pi_b$ be the distribution of $Z_b$ for $b\in\{0,1\}$ over $[A] \cup \{0\}$ and denote $\|\pi_1-\pi_0\|_1=\delta$. Lemma \ref{ter:lem:regularization} provides a distribution $\omega$ that has the same first $r-1$ moments as $\pi_0$ and $\pi_1$, is $\delta$-close to both $\pi_0$ and $\pi_1$, and satisfies $\omega(s)\ge 32^{-A}\delta$ for every $s\in [A] \cup \{0\}$. The distribution $\omega$ is used for the continuous hybrid approach: we will combine two paths $\pi_{b,u}=(1-u)\pi_b +u\omega$ for $b\in \{0,1\}$ that have the same moments up to order $r-1$ as $\pi_b$ and $\omega$. 

Denoting $F_b(u)=\E_{\mathbf z\sim\pi_{b,u}^{\otimes m}}[h(\mathbf z)]$, the goal is to bound $|F_1(0)-F_0(0)|\le |F_0(1)-F_0(0)| + |F_1(1)-F_1(0)|$. Similar to the previous theorem’s proof, we can use Lemma~\ref{lem:path-moments} to bound each term $|F_b(1)-F_b(0)|$ for $b \in \{0,1\}$. The second term of the bound obtained this way corresponds to the exceptional event $w>\frac{3N}{4}$, which can again be bounded using Hoeffding’s inequality. For the first term, instead of using the Taylor bound as before, we now use the Newton bound from Lemma~\ref{lem:newton} to bound the inside expression as follows:
$$\left|\sum_{s\in S}\dot\pi_{b,u}(s)
       \bigl(q_{\mathbf w}(s)-P_{\mathbf w}(s)\bigr)\right|
 \le\delta\sum_{\ell=r}^A\binom A\ell|d_\ell(\mathbf w)|.$$
 
Now we use the other main lemma of this proof: Lemma~\ref{ter:lem:average} provides an upper bound of $\frac{C^\ell T^{1+\alpha_\ell}}{N}(ma)^{-\alpha_\ell}$ on the expectation of $|d_\ell(\mathbf w)|$ where $\alpha_j=1-2^{1-j}$ and $a$ is a lower bound on the probabilities in $\pi_{b,u}$: this is why the lower bound on the probabilities in $\omega$ is crucial. %
Then, with little calculation (explained in full detail in \Cref{ter:sec:direct}), one can obtain the required upper bound
$$\left|\Pr_{f\sim\Pi_{0,N}}[\mathcal A^f=1]
       -\Pr_{f\sim\Pi_{1,N}}[\mathcal A^f=1]\right|
\le C^A\frac{T^{2-2^{1-r}}}{N^{1-2^{1-r}}}$$
 up to a small additive term bounded by $N\exp \left(-\frac{cN}{A^2}\right)$ due to the assumption $w\leq 3N/4$.
Thus, when $A$ and $r$ are fixed,
this quantity can only be large when the number of queries $T=\Omega(N^{1/2-1/(2(2^r-1))})$.

\paragraph{Obtaining the distribution $\omega$}
The proof of Lemma~\ref{ter:lem:regularization} first starts from $(\pi_0+\pi_1)/2$ and adds a small amount of mass uniformly, ensuring that every point receives positive mass.
However, this changes the total mass and the first $r-1$ moments.
Then, using Lagrange interpolation (Lemma~\ref{lem:grid-interpolation}), we adjust the masses at $r$ suitably chosen points to restore them. The interpolation bound controls this correction, preserving the required lower bound on every probability and the closeness to $\pi_0$ and $\pi_1$.

\paragraph{Newton coefficients of $q_{\mathbf w}$} %
Now we turn to proving \Cref{ter:lem:average}, and introduce some notation.
First, we write the input function $f:[N]\to[N]$ as a string $f\in[N]^N$, i.e., $f=(f(1),\dots,f(N))$. We partition the set of indices $[N]$ according to the value $f$ takes there, leading to ``blocks'' of indices, one for each value. The size of each block is the frequency of the corresponding value.
Let us take an $\ell\ge r$ such that $f$ contains at least one block of size $\ell$, and let $\rho$ denote the number of blocks of size $\ell$. Moreover, let $n_j$ denote one plus the number of blocks of size $j$ (for $j<\ell$), and $n_*=\min_{1\le j<\ell}n_j$. Also, let $p(f)=\Pr[\mathcal{A}^f=1]$ be the algorithm's acceptance probability on input $f$. 

We define the following distribution $\nu_j$ for each $j\le \ell$.
We first take a uniformly random partition of $[N]$ to block sizes prescribed by the frequency vector $\mathbf c$ (coming from the original input distribution), and mark a uniformly random block of size $\ell$. Now let us choose a size-$j$ subset of the marked block uniformly at random, call this subset $B$, and split the other $\ell-j$ positions into singletons. 
We assign distinct labels uniformly at random to the blocks, and define $f(i)$ as the label of the block containing $i$. Then we choose $\mathbf a=(a_i)_{i\in B}$ uniformly among all $j$-tuples of distinct labels from $[N]\setminus f([N])$. %
So $\nu_j$ is a distribution of tuples $(f,B,\mathbf a)$. One can see that the marginal distribution of $f$ according to $\nu_j$ is the same as $D_{\mathbf c(\mathbf{w},j)}$, thus $\E_{\nu_j}[p(f)]=q_\mathbf{w}(j)$. 

For a subset $U\subseteq B$, define $f^{U,\mathbf{a}}$ as the input one obtains from $f$ by turning $f(i)$ into $a_i$ for each $i\in U$. Based on this, let us define the block difference operator as 
\[
 D_B^{\mathbf a}g(f)=\sum_{U\subseteq B}(-1)^{|U|}g(f^{U,\mathbf a}),
\]
where $g$ is any (scalar or vector-valued) function on $[N]^N$. We usually omit $\mathbf{a}$ from the notation.
Take $j=\ell$ and $g=p$ (i.e., the acceptance probability). For each $U\subseteq B$, splitting it off leaves an $(\ell-|U|)$-block (and $|U|$ new  singletons). Because of uniformity, the expected acceptance probability for this term is $q_{\mathbf{w}}(\ell-|U|)$. Thus, using the definition of the Newton coefficients $d_\ell(\mathbf w)=\sum_{s=0}^\ell(-1)^{\ell-s}\binom\ell s q_{\mathbf w}(s)$ (for $0\le\ell\le A$), we obtain that (\Cref{lem:newton_block_difference}) 
$$\E_{\nu_\ell}[(D_Bp)(f)]
 =\sum_{s=0}^\ell(-1)^{\ell-s}\binom\ell s q_{\mathbf w}(s)
 =d_\ell(\mathbf w).$$

\paragraph{Bounding expected block differences} %
As we mentioned, \Cref{ter:lem:average}  gives an upper bound of $\frac{C^\ell T^{1+\alpha_\ell}}{N}(ma)^{-\alpha_\ell}$ on the expectation of $|d_{\ell}(\mathbf{w})|$. This is done by bounding each $|d_{\ell}(\mathbf{w})|$ and then averaging over $\mathbf{w}$ using the binomial bounds of Lemma~\ref{lem:inverse-binomial}. 
For bounding each $|d_{\ell}(\mathbf{w})|$, the special case when $\mathbf{w}\in\{0,1\}^{m-1}$ (which we call the singleton case) is handled separately in Lemma~\ref{ter:lem:singletons} simply by using the Newton bound (Lemma~\ref{lem:newton}). The other cases are treated in Lemma~\ref{ter:lem:fixed} by combining the last equation of the previous paragraph and the bound on the expected block differences $|\E_{\nu_\ell}[(D_Bp)(f)]|\le\frac{C^\ell T^{1+\alpha_\ell}}{\sqrt\rho\,n_1n_*^{\alpha_\ell-1/2}}$ (remember that $\alpha_j=1-2^{1-j}$) provided by Lemma~\ref{ter:lem:quantum}.
Finally, the bounds on the inverse binomial moments of \Cref{lem:inverse-binomial} turn the dependence on $\rho$ and $n_*$ to the required $(ma)^{-\alpha_\ell}$ factor.

\paragraph{Query-gain argument} %
Now we turn to proving Lemma~\ref{ter:lem:quantum}, which is the core of the proof of Theorem~\ref{thm:proportional-differences}. 
Our argument is similar to the query-gain argument in~\cite{Belovs26}, %
for which we only present here a high-level intuition.
Recall that the goal is to prove an upper bound on $|\E_{\nu_\ell}[(D_Bp)(f)]|$. 
Similarly to the standard adversary bound, the proof analyzes how the state of the $T$-query algorithm on input $f$ changes query by query. 
Let $\ket{\psi_t}$ be the state after query $t$ and $\ket{\psi'_t}$ the state before query $t+1$. 
We apply $D_B$ on the vector-valued function $f\mapsto \ket{\psi_t(f)}$.
Clearly, $\ket{\psi_0}$ is independent of the input, and $\ket{\psi_T}$ is closely related to the acceptance probability $p(f)$. For instance, if the accepting state is only $\ket{0}$, then $\langle 0|\psi_{T}(f)\rangle=1-2p(f)$. We deal with the general case in the main proof in \Cref{ter:sec:direct}.

Let us define difference vector $\ket{e_{B,t}}=D_B\ket{\psi_{t+1}}-\mathcal O_fD_B\ket{\psi'_t}$. 
One can telescope these differences to write $D_B\ket{\psi_T}$ (which is a linear function of $D_Bp$) as a rather simple combination of the $\ket{e_{B,t}}$ vectors. 
Lemma~\ref{ter:lem:geometry} combined with Claim~\ref{claim:claim} provides bound $\norm{\ket{e_{B,t}}}_{\nu_j}^2
 \le\frac{4j}{n_1}\frac{4^{2j}T^{2\alpha_{j}}}{n_1\,n_*^{2\alpha_{j}-1}}$.
Combining these facts, one can obtain that $|\E_{\nu_\ell}[(D_Bp)(f)]|$ can be upper bounded by the expression $\sum_{t=0}^{T-1}\norm{\ket{e_{B,t}}}_{\nu_\ell}/\sqrt{\rho} \le\frac{C^\ell T^{1+\alpha_\ell}}
              {\sqrt\rho\,n_1n_*^{\alpha_\ell-1/2}}$, thus finishing the proof.

\section{Problem setting and main results}\label{sec:setting}

\subsection{Construction of the input distributions}\label{sec:inputdistributions}

For an integer $N\ge1$, let $[N]=\{1,\ldots,N\}$.
We also usually write finite real vectors in bold face and their coordinates in normal font, such as
$\mathbf x=(x_1,\ldots,x_m)$. %

\paragraph{Frequency vector}
The \emph{frequency vector} of any given function $f:[N]\to[N]$ is
$\mathbf c(f)=(|f^{-1}(1)|,\ldots,|f^{-1}(N)|)$.
A vector $\mathbf c=(c_1,c_2,\ldots,c_N)\in\Z_{\ge0}^N$ is \emph{feasible} if
$\sum_{j\in[N]} c_j =N$, equivalently if $\mathbf c=\mathbf c(f)$ for some
$f:[N]\to[N]$. 
We call any such vector a \emph{feasible frequency vector}, or simply a feasible $\mathbf c$.

For a feasible (frequency vector) $\mathbf c$, let $\cD_{\mathbf c}$ be the uniform distribution on functions whose frequency vector is a permutation of $\mathbf c$.
To sample from $\cD_{\mathbf c}$, one can first uniformly permute the entries of $\mathbf c$ to obtain $\mathbf c'$, then choose $(f(1),\ldots,f(N))$ uniformly among all sequences in which each output label $j$ occurs exactly $c'_j$ times.

The following random variable describes the frequency of a uniformly chosen output label.
For a function $f$ drawn from $\cD_{\mathbf c}$, define its \emph{positive frequency statistics} $X=\abs{f^{-1}(Y)}$, where $Y$ is uniformly chosen from $f([N])$. Equivalently,
$$\Pr[X=a]=\frac{\abs{\{j:c_j=a\}}}{\abs{\{j:c_j\neq 0\}}}\quad(a\ge1).$$
Choosing instead $Y'$ uniformly from $[N]$ gives another random variable, the \emph{mean-one frequency statistics} $Z=\abs{f^{-1}(Y')}$. Since $\E[X]=N/\abs{f([N])}$, we have
\begin{equation}\label{eq:xtoz}
    \Pr[Z=a]=\frac{\Pr[X=a]}{\E[X]} \quad\text{(for $a\geq 1$)}, \qquad
\Pr[Z=0]=1-\frac{1}{\E[X]}.
\end{equation}
So the normalization simply corresponds to including unused output labels.
When it is clear from the context, we will refer to any of these statistics simply as frequency statistics.

\paragraph{From frequency statistics to functions}
We now construct a random function from prescribed positive frequency statistics $X$,
or equivalently its associated mean-one frequency statistics $Z$ as in \Cref{eq:xtoz}.
We use a randomized variant of the frequency construction of~\cite[Definition~5.4]{RRSS09}.
We sample $m$ multiplicities $\mathbf{z}=(z_1,\ldots,z_m)$ i.i.d. according to $Z$, each nonzero draw specifying the multiplicity of a distinct output label.
If their sum is $z\le N$, we add $N-z$ additional output labels, each occurring exactly once. Note that adding singleton labels changes the final frequency statistics.

\begin{definition}[Induced input distribution]\label{def:input-distributions}
Fix an integer $A\ge1$ and positive frequency statistics $X$
supported on $[A]$. Denote $\mu=\E[X]$, and define its associated
mean-one frequency statistics $Z$ by~\eqref{eq:xtoz}.
For $N\ge1$, let $\Pi_{N}$ be the distribution
of $f:[N]\to[N]$ obtained as follows:
\begin{enumerate}
\setlength{\itemsep}{0pt}
\item Draw $m=\lfloor N/2\rfloor$ independent multiplicities
      $\mathbf z=(z_1,\ldots,z_m)$ according to the distribution of $Z$; put $z=\sum_i z_i$.
\item If $z\le N$, form $\mathbf c=\mathbf c(\mathbf z)$ by taking the non-zero
      entries of $\mathbf z$, then adding $N-z$ entries equal to $1$, and finally zeros to length $N$.
      If $z>N$, set $\mathbf c=(1,\ldots,1)$.
\item Draw $f\sim\cD_{\mathbf c}$.
\end{enumerate}
\end{definition}
Observe that the vector $\mathbf c$ in step~2 is always feasible: its non-zero entries sum to $N$.

\begin{remark}\label{remark}
Let $S=\{0\}\cup[A]$, $m=\lfloor N/2\rfloor$. Let $\pi$ be the probability distribution of $Z$, that is $\pi(s)=\Pr[Z=s]$ for $s\in S$.
Then $\Pi_N$ is directly connected to $\pi^{\otimes m}$ as follows.
For $\mathbf z\in S^m$, let $\mathbf c(\mathbf z)$ be the frequency
vector obtained in step~2 of Definition~\ref{def:input-distributions}.
Then generating $f$ according to $\Pi_N$ is equivalent to generating ${\mathbf z}$ according to $\pi^{\otimes m}$ and then $f$ according to $\cD_{\mathbf c(\mathbf z)}$.    

In particular, for every algorithm $\mathcal{A}$ with query access to $f$,
we have
\begin{equation}\label{eq:remark}
 \Pr_{f\sim\Pi_N}[\mathcal A^f=1]
 =\E_{\mathbf z\sim\pi^{\otimes m}}[h(\mathbf z)],\quad\text{ with }
 h(\mathbf z)  =\Pr_{f\sim\cD_{\mathbf c(\mathbf z)}}[\mathcal A^f=1].
\end{equation}
\end{remark}

\begin{example}\label{ex:input-distribution}
Consider $A=3$ and $N=8$, and let $X$ equal $1$ or $3$, each with
probability $1/2$.  Then $\mu=2$ and
\[
 \Pr[Z=0]=\frac12,\qquad
 \Pr[Z=1]=\Pr[Z=3]=\frac14.
\]
To sample $f\sim\Pi_{8}$, draw four independent multiplicities from
this distribution.  If the draws are $(3,1,0,0)$, their sum is $s=4$, so we add
four output values of multiplicity one and obtain
\[
 \mathbf c=(3,1,1,1,1,1,0,0).
\]
Thus there is one output value of multiplicity three and five of multiplicity one.
One possible outcome after assigning labels and positions is
\[
 (f(1),\ldots,f(8))=(7,2,7,5,8,3,6,7).
\]
For comparison, the draw $(3,3,3,0)$ has $s=9>N$ and triggers the
fallback $\mathbf c=(1,\ldots,1)$, producing a uniformly random bijection.

The number $K$ of draws equal to $3$ has distribution $\operatorname{Bin}(4,1/4)$.
Values $K=0$ or $K\ge3$ produce a bijection, while $K=1,2$ produce one or two output values of multiplicity three, respectively.  Hence the
full input distribution is
$$ \Pi_{8}
 =\frac{47}{128}\cD_{(1,1,1,1,1,1,1,1)}
     +\frac{27}{64}\cD_{(3,1,1,1,1,1,0,0)}
    +\frac{27}{128}\cD_{(3,3,1,1,0,0,0,0)}.$$
\end{example}

\subsection{Main theorems}

Fix an integer $A\geq 1$, and two positive frequency statistics $X_0,X_1$ supported
on $[A]$. We define their respective induced input distributions $\Pi_{0,N}$ and $\Pi_{1,N}$ according to Definition~\ref{def:input-distributions}, and consider the following distinguishability problem.

A single function $f\sim\Pi_{b,N}$ is drawn before the algorithm starts.
The (quantum) algorithm receives standard (quantum) query access to this fixed function $f$, with $b$ hidden. Our goal is to bound the difference between its acceptance probabilities
under $\Pi_{0,N}$ and $\Pi_{1,N}$ in terms of the moments and the number of queries it makes.

The following proposition expresses exact moment matching of the mean-one frequency statistics as proportional moments of the positive frequency statistics.

Recall that, for $b\in\{0,1\}$ and $\mu_b=\E[X_b]$, we define $Z_b\in\{0\}\cup[A]$ by
\[
 \Pr[Z_b=a]=\frac{\Pr[X_b=a]}{\mu_b}\quad \text{(for $a\in[A]$)},
 \qquad \Pr[Z_b=0]=1-\frac1{\mu_b}.
\]
Since $\mu_b\ge1$, these probabilities are nonnegative and sum to one.
Equivalently, we keep an independent draw from $X_b$ with probability
$1/\mu_b$ and replace it by zero otherwise.
\begin{proposition}[Normalization of moments]\label{prop:normalization}
For $b\in\{0,1\}$ and every integer $j\ge1$,
\begin{equation}\label{eq:normalized-moments}
 \E [Z_b^j]=\frac{\E [X_b^j]}{\mu_b},
 \qquad \E [Z_b]=1.
\end{equation}
Consequently, for every integer $j\ge1$,
\[
 \E [Z_0^j]=\E [Z_1^j]
 \quad\Longleftrightarrow\quad
 \frac{\E [X_1^j]}{\E [X_0^j]}=\frac{\mu_1}{\mu_0}.
\]
\end{proposition}

\begin{proof}
The proof is direct using, for $j\ge1$.
\[
 \E [Z_b^j]=\sum_{a=1}^A a^j\frac{\Pr[X_b=a]}{\mu_b}
          =\frac{\E [X_b^j]}{\mu_b}.
\]
In particular, all denominators are non-zero since $X_b\ge 1$,
so cross-multiplication proves the equivalence.
The case $j=1$ provides $\E [Z_b]=1$.
\end{proof}

We now introduce our two main theorems, which share the same moment hypothesis.
Theorem~\ref{thm:proportional-polynomial} remains useful when the support
bound $A$ and the number $(r-1)$ of matched moments grow with $N$.
For fixed $A,r$, Theorem~\ref{thm:proportional-differences} gives a stronger
quantum query lower bound for constant distinguishing bias, at the cost
of an exponential factor $C^A$.

Theorem~\ref{thm:proportional-polynomial} is proved in Section~\ref{sec:polynomial-proof}
and Theorem~\ref{thm:proportional-differences} in Section~\ref{ter:sec:direct}.
The constants $c,C$ can be chosen the same in both theorems.

\begin{theorem}\label{thm:proportional-polynomial}
Fix an integer $A\ge1$ and positive frequency statistics $X_0,X_1$
supported on $[A]$. 
Define $\Pi_{0,N},\Pi_{1,N}$ according to Definition~\ref{def:input-distributions} applied to $X_0$ and $X_1$.
Suppose that they satisfy the following, for an integer $r\ge2$:
\begin{equation}\label{eq:proportional}
 \frac{\E [X_1]}{\E [X_0]}
 =\frac{\E [X_1^2]}{\E [X_0^2]}
 =\cdots
 =\frac{\E [X_1^{r-1}]}{\E [X_0^{r-1}]}.
\end{equation}
There are absolute constants $c,C>0$ such that, for integers
$N\ge CA$ and $0\le T\le c\sqrt N$, every $T$-query quantum algorithm
$\mathcal A$ %
satisfies
\begin{equation}\label{eq:proportional-polynomial}
\left|\Pr_{f\sim\Pi_{0,N}}[\mathcal A^f=1]
       -\Pr_{f\sim\Pi_{1,N}}[\mathcal A^f=1]\right|
\le \frac NA\left(\frac{CAT^2}{r^2N}\right)^r
 +N\exp \left(-\frac{cN}{A^2}\right).
\end{equation}
In particular, if $A=O(r^2)$ and $A=o(\sqrt{N/\log N})$,
this difference is $o(1)$ whenever
$T=o(N^{1/2-1/(2r)})$ as $N\to\infty$.
Each probability includes the initial draw of $f$ and the algorithm's
measurement outcomes.
\end{theorem}

\begin{theorem}\label{thm:proportional-differences}
Fix an integer $A\ge1$ and positive frequency statistics $X_0,X_1$
supported on $[A]$. 
Define $\Pi_{0,N},\Pi_{1,N}$ according to Definition~\ref{def:input-distributions} applied to $X_0$ and $X_1$.
Suppose that they satisfy the following, for an integer $r\ge2$:
\[
 \frac{\E [X_1]}{\E [X_0]}
 =\frac{\E [X_1^2]}{\E [X_0^2]}
 =\cdots
 =\frac{\E [X_1^{r-1}]}{\E [X_0^{r-1}]}.
\]
There are absolute constants $c,C>0$ such that, for integers
$N\ge CA$ and $0\le T\le c\sqrt N$, every $T$-query quantum algorithm
$\mathcal A$ %
satisfies
\begin{equation}\label{eq:proportional-differences}
\left|\Pr_{f\sim\Pi_{0,N}}[\mathcal A^f=1]
       -\Pr_{f\sim\Pi_{1,N}}[\mathcal A^f=1]\right|
\le C^A\frac{T^{2-2^{1-r}}}{N^{1-2^{1-r}}}
 +N\exp \left(-\frac{cN}{A^2}\right).
\end{equation}
In particular, for fixed $A,r$,
this difference is $o(1)$ whenever
$T=o(N^{1/2-1/(2(2^r-1))})$ as $N\to\infty$.
Each probability includes the initial draw of $f$ and the algorithm's
measurement outcomes.
\end{theorem}

\paragraph{Remark on the case $r=2$.}
This base case  also appears in~\cite[Definition~5.4, Theorem~5.6 and
Corollary~5.7]{RRSS09}. When $r=2$, the proportional-moment condition is vacuous, so both theorems apply
directly. Consider $A=2$, $X_0\equiv1$, and $X_1\equiv2$.
Then $\Pi_{0,N}$ is uniform on bijections, whereas $\Pi_{1,N}$ has
$K\sim\operatorname{Bin}(\lfloor N/2\rfloor,1/2)$ output values of
multiplicity $2$, completed with singletons. With probability
$1-e^{-\Omega(N)}$, at least $N/8$ such values occur, so the function
is at distance at least $1/8$ from injectivity.
This is element distinctness testing: injective versus far from injective;
the second distribution need not be two-to-one.
Theorem~\ref{thm:proportional-polynomial} gives an $\Omega(N^{1/4})$
quantum query lower bound, while Theorem~\ref{thm:proportional-differences}
recovers the known $\Omega(N^{1/3})$ bound for testing injectivity.
The latter follows from the lower bound on the collision problem \cite{aaronson2004quantum} by noticing that a two-to-one function is a special case of a far from injective function.

\section{Moment matching: framework and first bound}\label{sec:polynomial-proof}
We begin by introducing the general tools that will be used in the proofs of our two main theorems, and then prove the first one, namely Theorem~\ref{thm:proportional-polynomial}.

\subsection{Polynomial and probabilistic tools}

\subsubsection{Symmetrization}

We use the frequency-vector notation of
Subsection~\ref{sec:inputdistributions}. The following lemma establishes the connection between quantum query complexity and the moment-matching technique used in our proof. It shows that, after averaging over random permutations of the domain and range, the acceptance probability of any quantum algorithm making at most $T$ queries can be expressed as a symmetric polynomial of degree at most $2T$ in the output frequencies. 

Moreover, its representation in terms of power sums, together with the weighted-degree restriction, allows us to obtain a low-degree univariate polynomial when varying a single multiplicity while keeping others fixed. This polynomial structure is crucial for applying moment matching to cancel the lower-order terms and using polynomial inequalities to bound the remaining contribution.

\begin{lemma}[Symmetrization]\label{lem:symmetrization}
For every quantum algorithm $\mathcal A$ making at most $T$ queries,
there is a symmetric polynomial $P\in\R[x_1,\ldots,x_N]$ of degree
at most $d=2T$ such that, for every feasible $\mathbf c$,
\[
 P(\mathbf c)=\Pr_{f\sim\cD_{\mathbf c}}[\mathcal A^f=1]\in[0,1].
\]
If $d\ge1$, one can write
$P(\mathbf x)=F(p_1(\mathbf x),\ldots,p_d(\mathbf x))$,
where $p_j(x)=\sum_{i=1}^N x_i^j$ is the $j$-th power-sum symmetric polynomial,
 $F\in\R[y_1,\ldots,y_d]$ and every monomial
$y_1^{\alpha_1}\cdots y_d^{\alpha_d}$ of $F$
satisfies $\sum_{j=1}^d j\alpha_j\le d$.
\end{lemma}
\begin{proof}
First, following~\cite{BBCMW01}, %
the acceptance probability $\Pr[\mathcal A^f=1]$ is a polynomial
of degree at most $d$ in the indicators
$\delta_{i,a}(f)=\mathbf1[f(i)=a]$.

For an input with frequency vector $\mathbf c$, averaging over
independent uniform domain and range permutations induces
$\cD_{\mathbf c}$.
On valid encodings of $f$, %
each monomial either vanishes or is the indicator of
a partial function.  If it fixes $q\leq d$ domain points and uses $\ell$ distinct output
values $a_1,\ldots,a_\ell$ with multiplicities $q_1,\ldots,q_\ell$, its average is
\[
 \frac{1}{\falling{N}{q}\falling{N}{\ell}}
 \sum_{\substack{a_1,\ldots,a_\ell\in[N]\\\text{distinct}}}
 \prod_{s=1}^\ell\falling{\mathbf{c}_{a_s}}{q_s},
 \quad \text{where } \falling{x}{k}=x(x-1)\cdots(x-k+1), \text{ and } \falling{x}{0}=1.
\]

This is a symmetric polynomial of degree at most $\sum_s q_s=q\leq d$, proving the assertion about $P$.
The fundamental theorem of symmetric polynomials and Newton's identities
express each homogeneous component of $P$ in power sums of the same
weighted degree~\cite{Macdonald95}, giving $F$.
\end{proof}

\subsubsection{Approximation and moment bounds}\label{sec:approx-moments-bounds}
We begin by stating some polynomial tools that will be used in the proofs of Theorems~\ref{thm:proportional-polynomial} and~\ref{thm:proportional-differences}. The first provides a bound on the Taylor remainder of a low-degree polynomial whose values are bounded on an integer interval, while the second gives a Newton expansion and a corresponding remainder estimate. These results will allow us to control the higher-order terms that remain after moment matching. For completeness, we provide their proofs in %
Appendix~\ref{appendix:proof of aux results}.

\begin{restatable}[Taylor bound]{lemma}{taylorlemma}\label{lem:taylor}
Let $H\ge1$ and $d\ge0$ be integers with $d^2\le H/2$.
Let $q(z)=\sum_{\ell=0}^{d}a_\ell z^\ell$ be any polynomial of degree at most $d$ such that $|q(z)|\le1$
for $z\in \{0\}\cup[H]=\{0,1,\ldots,H\}$. 
For an integer $r\ge1$, let $Q_r=\sum_{\ell=0}^{\min(r-1,r)}a_\ell z^\ell$.
Then for every real $z\in[0,H]$,
\begin{equation}\label{eq:taylor}
 \left|q(z)-Q_r(z)\right|
 \le\frac{2}{(2r)!}\left(\frac{4d^2}{H}\right)^r z^r.
\end{equation}
\end{restatable}

\begin{restatable}[Newton bound]{lemma}{newtonlemma}\label{lem:newton} 
Let $H\ge1$ and $d\ge0$ be  integers, and let $q(z)$ be any polynomial of degree at most $d$ such that $|q(z)|\le1$ for $z\in \{0\}\cup[H]$. 
Its \emph{Newton coefficients} are
\[
 d_\ell=\sum_{s=0}^\ell(-1)^{\ell-s}\binom\ell s q(s)
 \qquad(0\le\ell\le H).
\]
For an integer $1\le r\le H+1$, let
$Q_r(z)=\sum_{\ell=0}^{r-1}d_\ell\binom z\ell$.
At every $z\in\{0\}\cup[H]$,
\[
  q(z)-Q_r(z)=\sum_{\ell=r}^H d_\ell\binom z\ell=\sum_{\ell=r}^z d_\ell\binom z\ell,
\]
where empty sums are understood to be zero. %

Assume moreover that $d^2\le H/2$. Then, for every integer $1\le r\le H+1$ and $z\in\{0\}\cup[H]$,
\begin{equation}\label{eq:newton-remainder}
 |q(z)-Q_r(z)|
 \le\frac{2}{(2r)!}\left(\frac{4d^2}{H}\right)^r z^r.
\end{equation}
\end{restatable}

We first establish a basic consequence of moment matching. The following lemma shows that when two probability distributions have matching moments up to a given order, the corresponding lower-order terms vanish in the difference of expectations of a polynomial. This cancellation holds in both the monomial and Newton bases and will be used to control the differences in acceptance probabilities.

\begin{lemma}[Moment cancellation]\label{lem:moment-cancellation}
Let $A,r\ge1$ be integers, and let $\pi,\nu$ be probability distributions
on $\{0\}\cup[A]$ with matching moments through order $r-1$, that is,
\[
 \sum_{s=0}^{A}\nu(s)s^k =  \sum_{s=0}^{A}\pi(s)s^k
 \qquad(k=1,\ldots,r-1).
\]
Choose one of the two bases
$\phi_\ell(z)=z^\ell$, or $\phi_\ell(z)=\binom z\ell$, for $\ell\ge0$.
Suppose $q(s)=\sum_{\ell=0}^D a_\ell\phi_\ell(s)$ on
$\{0\}\cup[A]$, for an integer $D\ge0$. Then
\begin{equation}\label{eq:moment-cancellation}
 \begin{aligned}
 & \sum_{s=0}^A(\nu(s)-\pi(s))q(s)
   =\sum_{\ell=r}^D\eta_\ell a_\ell,\\
\textit{ where }\quad &   \eta_\ell=\sum_{s=0}^A(\nu(s)-\pi(s))\phi_\ell(s),
 \text{ and } 
 |\eta_\ell|\le\norm{\nu-\pi}_1\phi_\ell(A). 
 \end{aligned}
\end{equation}
\end{lemma}
\begin{proof}
The moment identities also hold for $k=0$, since both distributions have total mass $1$. For either basis, $\phi_\ell$ is a polynomial of degree $\ell$. Thus, for $\ell<r$, writing
$\phi_\ell(s)=\sum_{k=0}^{\ell}b_k s^k$ gives
\[
 \eta_\ell
 =\sum_{k=0}^{\ell}b_k
   \sum_{s=0}^A(\nu(s)-\pi(s))s^k
 =0.
\]
Expanding $q$ and interchanging the finite sums therefore yields
\[
 \sum_{s=0}^A(\nu(s)-\pi(s))q(s)
 =\sum_{\ell=0}^D a_\ell\eta_\ell
 =\sum_{\ell=r}^D a_\ell\eta_\ell.
\]
Finally, for either basis,
$0\le\phi_\ell(s)\le\phi_\ell(A)$ on $\{0\}\cup[A]$, so
\[
 |\eta_\ell|
 \le\sum_{s=0}^A|\nu(s)-\pi(s)|\phi_\ell(s)
 \le\norm{\nu-\pi}_1\phi_\ell(A).
\]
\end{proof}

We next recall the Lagrange polynomial interpolation result on integer intervals. For completeness, we provide a proof in Appendix~\ref{appendix:proof of aux results}. 
\begin{restatable}[Lagrange interpolation]{lemma}{gridinterpolationlemma}\label{lem:grid-interpolation}
Let $A\ge2$ and $0\le d\le A$ be integers, and let
$s_0<\cdots<s_d$ belong to $\{0\}\cup[A]$. Define
\[
 L_i(x)=\prod_{\substack{0\le h\le d\\h\ne i}}
             \frac{x-s_h}{s_i-s_h}\qquad(0\le i\le d).
\]
Every polynomial $P$ of degree at most $d$ satisfies $P(x)=\sum_{i=0}^d P(s_i)L_i(x)$.
Moreover, at every integer $t\in\{0\}\cup[A]$, $\sum_{i=0}^d|L_i(t)|\le4^{A-1}$.
\end{restatable}

The following lemma provides bounds on inverse moments of binomial random variables. These estimates will be used to control the inverse block-count factors that arise when averaging the bounds on the Newton coefficients, required later in the proof of \Cref{thm:proportional-differences}. Importantly, this result does not require independence between the random variables.

\begin{lemma}[Inverse binomial moments]\label{lem:inverse-binomial}
Let $m\ge1$ be an integer, let $a,b\in(0,1]$, and let $Y,Z$ have marginal distributions
$\operatorname{Bin}(m-1,a)$ and $\operatorname{Bin}(m-1,b)$.
They need not be independent. Then
\[
 \E [(1+Y)^{-1}]=\frac{1-(1-a)^m}{ma}\le\frac1{ma}.
\]
For $\beta,\gamma\ge0$ with $\beta+\gamma\le1$,
\[
 \E\!\left[(1+Y)^{-\beta}(1+Z)^{-\gamma}\right]
 \le(ma)^{-\beta}(mb)^{-\gamma}.
\]
\end{lemma}
\begin{proof}
Since $Y\sim\operatorname{Bin}(m-1,a)$, we have
\[
 \E[(1+Y)^{-1}]
 =\int_0^1 \E[t^Y]\,dt
 =\int_0^1(1-a+at)^{m-1}\,dt
 =\frac{1-(1-a)^m}{ma}
 \le\frac1{ma}.
\]
The same bound holds for $Z$ with $b$ in place of $a$.
H\"older's inequality with weights $\beta,\gamma,1-\beta-\gamma$
(the third factor being $1$) then gives
\[
 \E[(1+Y)^{-\beta}(1+Z)^{-\gamma}]
 \le \E[(1+Y)^{-1}]^\beta
      \E[(1+Z)^{-1}]^\gamma
 \le (ma)^{-\beta}(mb)^{-\gamma}.
\]
\end{proof}

\subsection{Hybrid arguments and  moment matching}\label{sec:product-comparison}

As discussed in the overview, interpolation between multiplicity distributions turns the global comparison into a one-coordinate calculation. This is a reminiscence of the quantum hybrid argument in~\cite{BBBV97}.
The two lemmas of this section make this reduction explicit. 

We start by defining a new notion of continuous hybrid argument.
Then we explain why it is generalizing the usual one. Finally, we state a lemma that will be use to exploit the moment-matching condition.

In this full subsection, fix integers $A,m\ge1$, and let $S=\{0\}\cup[A]$. 
They will be used as follows: \(A\) is the largest possible sampled multiplicity, \(m\) is the number of multiplicities we sample and \(S\) is the set of possible values of one multiplicity.

\subsubsection{Continuous hybrid argument}

Let $h:S^m\to[0,1]$ be invariant under permutations of its coordinates.
Symmetry lets us move any chosen coordinate to the last position. 
For this reason, for $\mathbf z=(\mathbf w,s)\in S^m$, with  $\mathbf w\in S^{m-1}$ and $s\in S$,
we define $q_{\mathbf w}(s)=h(\mathbf w,s)=h(\mathbf z)$.

The following derivative identity in Lemma~\ref{lem:product-comparison}
provides a continuous version of the usual hybrid bound. We will elaborate on this after presenting its proof. Its final inequality additionally
takes the absolute value inside the expectation over $\mathbf w$.
\begin{lemma}[Continuous hybrid argument]\label{lem:product-comparison}
Let $h:S^m\to[0,1]$ be invariant under permutations of its coordinates.
Define $q_{\mathbf w}(s)=h(\mathbf w,s)$, for $\mathbf w\in S^{m-1}$.
Let $(\pi_u)_{0\le u\le1}$ be a continuous, piecewise $C^1$ path of probability distributions on $S$, and denote $F(u)=\E_{\pi_u^{\otimes m}}[h]$. 
At every differentiability point of the path,
with  $\dot\pi_u(s)=\frac{d}{du}\pi_u(s)$, the following holds:
\begin{equation}\label{eq:product-derivative}
 F'(u)=m \E_{\mathbf w\sim\pi_u^{\otimes(m-1)}}
                   \left[\sum_{s\in S}\dot\pi_u(s)q_{\mathbf w}(s)\right].
\end{equation}
Consequently,
\begin{equation}\label{eq:product-integral}
 |F(1)-F(0)|\le m\int_0^1
 \E_{\mathbf w\sim\pi_u^{\otimes(m-1)}}
 \left[\left|\sum_{s\in S}\dot\pi_u(s)q_{\mathbf w}(s)\right|\right]\,du.
\end{equation}
\end{lemma}

\begin{proof}
Writing the expectation as a finite sum and differentiating each product term gives us
\begin{eqnarray*}
 F'(u)&=&
  \frac{d}{du}\left(\sum_{\mathbf z\in S^m}h(\mathbf z)
                         \prod_{j=1}^m\pi_u(z_j)   \right)\\
 &=&\sum_{i=1}^m\sum_{\mathbf z\in S^m}
       h(\mathbf z)\dot\pi_u(z_i)
                     \prod_{\substack{1\le j\le m\\j\ne i}}\pi_u(z_j).
\end{eqnarray*}
For each fixed $i$, let us set $s=z_i$ and
$\mathbf w=(z_1,\ldots,z_{i-1},z_{i+1},\ldots,z_m)$.
Symmetry gives $h(\mathbf z)=h(\mathbf w,s)=q_{\mathbf w}(s)$.
Hence the inner sum is
\[
 \sum_{\mathbf w\in S^{m-1}}
       \left(\prod_{j=1}^{m-1}\pi_u(w_j)\right)
       \sum_{s\in S}\dot\pi_u(s)q_{\mathbf w}(s)
 =\E_{\mathbf w\sim\pi_u^{\otimes(m-1)}}\,
       \left[\sum_{s\in S}\dot\pi_u(s)q_{\mathbf w}(s)\right].
\]
This expression is independent of $i$, so summing over $i$ proves
\eqref{eq:product-derivative}.

Finally, $F$ is continuous and piecewise $C^1$, so
$ |F(1)-F(0)|\le\int_0^1|F'(u)|\,du$. Therefore we can conclude by
using
$|\E [X]|\le\E [|X|]$. %
\end{proof}

\subsubsection{Connection to standard hybrid arguments}

In our applications, these coordinates are multiplicity draws: $s$ is the value of one draw and $\mathbf w$ contains the other $m-1$ values.
More explicitly, \(h(\mathbf z)\) is the mean of the algorithm’s acceptance probability after fixing the multiplicity vector  \(\mathbf z\) and averaging over the subsequent random functions in $\cD_{\mathbf c(\mathbf z)}$. So $h$ is indeed symmetric. 
We remind that, according to \Cref{eq:remark} of Remark~\ref{remark}, 
if each coordinate of $\mathbf z$ is sampled i.i.d. according to the same distribution $\pi$ on $S$, then 
$$ \Pr_{f\sim\Pi_{N}}[\mathcal A^f=1]
 =\E_{\mathbf z\sim\pi^{\otimes m}}[h(\mathbf z)].$$ 

\paragraph{Standard hybrid argument}
We now describe the usual hybrid argument to analyze the success probability of any given algorithm against two different input distributions.
For two probability distributions $\pi_0,\pi_1$ on $S$, define the standard hybrids as
\[
 H_i=\E_{\mathbf z\sim \pi_1^{\otimes i}\otimes\pi_0^{\otimes(m-i)}}[h(\mathbf z)]
 \quad(0\le i\le m).
\]
We are interested in the quantity $H_m-H_0$, i.e., the difference of the expected acceptance probabilities when the whole input is sampled according to $\pi_0$ or $\pi_1$.
The usual hybrid changes one coordinate at each step and telescopes
$H_m-H_0=\sum_{i=0}^{m-1}(H_{i+1}-H_i)$.
By symmetry, each increment is
\[
 H_{i+1}-H_i
 =\E_{\mathbf w\sim\pi_1^{\otimes i}\otimes\pi_0^{\otimes(m-1-i)}}
 \left[\sum_{s\in S}(\pi_1(s)-\pi_0(s))q_{\mathbf w}(s)\right].
\]

\paragraph{Continuous hybrid argument}
We now give a continuous version of this argument in Lemma~\ref{lem:product-comparison}.
Let $(\pi_u)_{0\le u\le1}$ be a continuous, piecewise $C^1$ path of
probability distributions on $S$ joining $\pi_0$ to $\pi_1$, and let
\[
 F(u)=\E_{\mathbf z\sim\pi_u^{\otimes m}}[h(\mathbf z)],\qquad
 \dot\pi_u(s)=\frac{d}{du}\pi_u(s),\qquad
 \norm{\dot\pi_u}_1=\sum_{s\in S}|\dot\pi_u(s)|.
\]

Here $F(0)=H_0$ and $F(1)=H_m$.
To see the connection with the usual hybrid argument, consider the special case of the linear path $\pi_u=(1-u)\pi_0+u\pi_1$. Then symmetry gives
\[
 F(u)=\sum_{i=0}^{m}\binom mi u^i(1-u)^{m-i}H_i.
\]
Differentiating with respect to $u$ yields 
\begin{eqnarray*}
 F'(u)& =&\sum_{i=1}^{m}i\binom mi
   u^{i-1}(1-u)^{m-i}H_i
-\sum_{i=0}^{m-1}(m-i)\binom mi
   u^i(1-u)^{m-1-i}H_i.\\ 
 &=&m\sum_{i=0}^{m-1}\binom{m-1}{i}
 u^i(1-u)^{m-1-i}(H_{i+1}-H_i).
\end{eqnarray*}
The weights are nonnegative and each integrates to $1$, so
\[
 |F(1)-F(0)| = |H_m-H_0|
 \le\int_0^1|F'(u)|\,du
 \le\sum_{i=0}^{m-1}|H_{i+1}-H_i|.
\]

Thus the derivative identity in Lemma~\ref{lem:product-comparison}
recovers the standard hybrid bound.

\paragraph{Final remark} 
The continuous formulation in \Cref{lem:product-comparison} is useful in Section~\ref{ter:sec:direct}
because the remaining draws are independent and have the same distribution:
the count $Y_s=|\{j:w_j=s\}|$ has distribution
$\operatorname{Bin}(m-1,\pi_u(s))$.
This permits the direct use of Lemma~\ref{lem:inverse-binomial} in
Lemma~\ref{ter:lem:average}.
A fixed discrete hybrid generally has two different coordinate distributions
and would require corresponding count estimates.
For the polynomial proof in Section~\ref{sec:polynomial-proof},
the standard discrete hybrid would give the same bound.

\subsubsection{Moment matching}

Moment matching allows us to replace \(q_{\mathbf w}\) by its approximation error \(q_{\mathbf w}-P_{\mathbf w}\) in the derivative formula of Lemma~\ref{lem:product-comparison}.
We state this below in a lemma that will provide us  a common comparison bound for
Theorems~\ref{thm:proportional-polynomial} and~\ref{thm:proportional-differences}. 
It also bounds the contribution of exceptional choices of $\mathbf w$
for which we do not control the approximation error.
\begin{lemma} %
\label{lem:path-moments}
Let $h:S^m\to[0,1]$ be invariant under permutations of its coordinates.
Define $q_{\mathbf w}(s)=h(\mathbf w,s)$, for $\mathbf w\in S^{m-1}$.
Let $(\pi_u)_{0\le u\le1}$ be a continuous, piecewise $C^1$ path of
probability distributions on $S$ whose moments through order $r-1$
are constant in $u$, for an integer $r\ge1$.
Let us denote $F(u)=\E_{\pi_u^{\otimes m}}\,[h]$.
For each $\mathbf w\in S^{m-1}$, let $P_{\mathbf w}$ be a polynomial
of degree below $r$.
For every $\mathcal G\subseteq S^{m-1}$,
\begin{equation}\label{eq:product-comparison}
 \begin{aligned}
 |F(1)-F(0)|
 &\le m\int_0^1
 \E_{\mathbf w\sim\pi_u^{\otimes(m-1)}}\!
 \left[\mathbf1_{\mathcal G}
 \left|\sum_{s\in S}\dot\pi_u(s)
       \bigl(q_{\mathbf w}(s)-P_{\mathbf w}(s)\bigr)\right|\right]\,du\\
 &\quad+m\int_0^1\norm{\dot\pi_u}_1
 \Pr_{\mathbf w\sim\pi_u^{\otimes(m-1)}}[\mathbf w\notin\mathcal G]\,du.
 \end{aligned}
\end{equation}
\end{lemma}
\begin{proof}
Fix $\mathbf w$. Then for two values $u,v\in[0,1]$, the two distributions $\pi_u,\pi_v$ 
have matching moments through order $r-1$. Since
$\deg P_{\mathbf w}<r$, Lemma~\ref{lem:moment-cancellation} gives
$\sum_{s\in S}(\pi_u(s)-\pi_v(s))P_{\mathbf w}(s)=0$.
Thus $\sum_{s\in S}\pi_u(s)P_{\mathbf w}(s)$ is constant in $u$.

Therefore, at every point where it is differentiable,
\[
 0=\frac{d}{du}\left(\sum_{s\in S}\pi_u(s)P_{\mathbf w}(s)\right)
  =\sum_{s\in S}\dot\pi_u(s)P_{\mathbf w}(s).
\]
Consequently,
\begin{equation}\label{eq:path-cancellation}
 \sum_{s\in S}\dot\pi_u(s)q_{\mathbf w}(s)
 =\sum_{s\in S}\dot\pi_u(s)
                  \bigl(q_{\mathbf w}(s)-P_{\mathbf w}(s)\bigr).
\end{equation}
Equation~\eqref{eq:product-integral} of Lemma~\ref{lem:product-comparison},
split over $\mathcal G$ and its complement, gives
\[
 \begin{aligned}
 |F(1)-F(0)|
 &\le m\int_0^1
 \E_{\mathbf w\sim\pi_u^{\otimes(m-1)}}\!
 \left[\mathbf1_{\mathcal G}
       \left|\sum_{s\in S}\dot\pi_u(s)q_{\mathbf w}(s)\right|\right]\,du\\
 &\quad+m\int_0^1
 \E_{\mathbf w\sim\pi_u^{\otimes(m-1)}}\!
 \left[\mathbf1_{\mathcal G^c}
       \left|\sum_{s\in S}\dot\pi_u(s)q_{\mathbf w}(s)\right|\right]\,du.
 \end{aligned}
\]
We apply~\eqref{eq:path-cancellation} in the first term.
For the second term, $0\le q_{\mathbf w}\le1$ gives
\[
 \left|\sum_{s\in S}\dot\pi_u(s)q_{\mathbf w}(s)\right|
 \le\sum_{s\in S}|\dot\pi_u(s)|q_{\mathbf w}(s)
 \le\norm{\dot\pi_u}_1.
\]
These bounds prove~\eqref{eq:product-comparison}, completing the proof.
\end{proof}

\subsection{Proof of Theorem~\ref{thm:proportional-polynomial}}

We are now almost ready to prove our first main theorem. This full subsection is devoted to this.
Let us fix the parameters of Theorem~\ref{thm:proportional-polynomial} and let $\mathcal A$ be a quantum algorithm making at most $T$ queries to $f:[N]\to[N]$.
The case $T=0$ is immediate. 

Assume
$N\ge\max\{8,4A\}$ and $T\ge1$ with $T^2\le N/64$.
Put $S=\{0\}\cup[A]$, $m=\lfloor N/2\rfloor$, and
$H=\lfloor N/4\rfloor$, and let $\pi_b$ be the distribution
of $Z_b$ for $b\in\{0,1\}$.
For $\mathbf z\in S^m$, let $\mathbf c(\mathbf z)$ be the frequency vector obtained in step~2 of Definition~\ref{def:input-distributions}, and define
\[
 h(\mathbf z)
 =\Pr_{f\sim\cD_{\mathbf c(\mathbf z)}}[\mathcal A^f=1].
\] %
Note that with this symmetrization, $h$ is not yet necessarily a polynomial in $\mathbf{z}$.

For $\mathbf w\in S^{m-1}$, let $w=\sum_iw_i$ and
$q_{\mathbf w}(s)=h(\mathbf w,s)$.
We first establish the following polynomial approximation.

\begin{lemma}[Taylor approximation]
\label{lem:fixed-multiplicities}
For every $\mathbf w\in S^{m-1}$ with $w\le3N/4$,
the function $q_{\mathbf w}$ admits a $2T$-degree polynomial extension,
also denoted $q_{\mathbf w}$, such that
\[
 q_{\mathbf w}(s)=\sum_{j=0}^{2T}a_{\mathbf w,j}s^j
 \qquad(s\in S),
\]
with real coefficients and satisfying
$|q_{\mathbf w}(z)|\le1$, for every $z\in\{0\}\cup[H]$.
Then every mean-one random variable $Z$ supported on $S$ satisfies
\begin{equation}\label{eq:local-taylor-residual}
\begin{aligned}
& \E_Z[|q_{\mathbf w}(Z)-Q_{\mathbf w}(Z)|]
 \le\frac{2A^{r-1}}{(2r)!}
       \left(\frac{128T^2}{N}\right)^r\\
\text{ where }\quad& Q_{\mathbf w}(z)=
\sum_{j=0}^{\min\{r-1,2T\}}a_{\mathbf w,j}z^j.
\end{aligned}
\end{equation}
\end{lemma}

\begin{proof}
Let us fix $\mathbf w$ with $w\le3N/4$ and set $d=2T$.
The parameter conditions give us $H\ge N/8$, $A\le H$, and $d^2\le H/2$.

As in step~2 of Definition~\ref{def:input-distributions}, for each
integer $0\le z\le H$, we form $\mathbf c(z)$ by taking the non-zero
entries of $(\mathbf w,z)$, adding $N-w-z$ entries equal to one,
and padding with zeros, if required, to make it of length $N$.
This vector is feasible since $w+z\le N$.
We extend $q_{\mathbf w}$ to these integers by
\[
 q_{\mathbf w}(z)
 =\Pr_{f\sim\cD_{\mathbf c(z)}}[\mathcal A^f=1]\in[0,1].
\]
For $z\in S$, we have $w+z\le N$, so this is exactly the
completion used to define $h(\mathbf w,z)$.

By Lemma~\ref{lem:symmetrization}, we know that there is a symmetric polynomial $P$
of degree at most $d$ such that
\[
 P(\mathbf c)=\Pr_{f\sim\cD_{\mathbf c}}[\mathcal A^f=1]
 \quad\text{for every feasible }\mathbf c.
\]
Let us write $P(\mathbf x)=G(p_1(\mathbf x),\ldots,p_d(\mathbf x))$,
where $G$ has weighted degree at most $d$.
For $1\le j\le d$, we have
$p_j(\mathbf c(z))=p_j(\mathbf w)+N-w-z+z^j$, where
the terms come from the fixed multiplicities, the completion singletons,
and the added multiplicity $z$.
Consequently $q_{\mathbf w}(z)$ is indeed a degree-$2T$ polynomial in $\mathbf{z}$: 
\[
 q_{\mathbf w}(z)
 =G\!\left(\bigl(p_j(\mathbf w)+(N-w-z)+z^j\bigr)_{j=1}^d\right)
 =\sum_{j=0}^{2T}a_{\mathbf w,j}z^j
\]
extends the integer values and has degree at most $d$.

Since $Q_{\mathbf w}$ is its  truncation below degree $r$,
Lemma~\ref{lem:taylor} gives
\[
 \E[|q_{\mathbf w}(Z)-Q_{\mathbf w}(Z)|]
 \le\frac{2}{(2r)!}\left(\frac{4d^2}{H}\right)^r\E[Z^r]
 \le\frac{2A^{r-1}}{(2r)!}\left(\frac{128T^2}{N}\right)^r,
\]
using the fact that $\E[Z^r]\le A^{r-1}\E[Z]=A^{r-1}$, $H\ge N/8$, and $d=2T$.
\end{proof}

Finally, we are ready to prove our first main theorem.
\begin{proof}[End of proof of Theorem~\ref{thm:proportional-polynomial}]
By Proposition~\ref{prop:normalization} and the moment matching
condition~\eqref{eq:proportional}, the distributions $\pi_0,\pi_1$ of
$Z_0,Z_1$ have mean one and matching moments through order $r-1$.
Let us apply on $S$ the linear path of probability distributions $\pi_u=(1-u)\pi_0+u\pi_1$ for $0\le u\le1$.
Its moments through order $r-1$ are constant, since
\[
 \E_{s\sim\pi_u}[s^j]
 =(1-u)\E[Z_0^j]+u\E[Z_1^j]
 =\E[Z_0^j]
 \qquad(1\le j\le r-1).
\]

Let $F(u)=\E_{\mathbf z\sim\pi_u^{\otimes m}}[h(\mathbf z)]$, so
$F(b)=\Pr_{f\sim\Pi_{b,N}}[\mathcal A^f=1]$ for $b=0,1$.
Apply Lemma~\ref{lem:path-moments} with
\[
 \mathcal G=\{\mathbf w\in S^{m-1}:w\le3N/4\},\qquad
 P_{\mathbf w}=\begin{cases}
 Q_{\mathbf w},&\mathbf w\in\mathcal G,\\
 0,&\mathbf w\notin\mathcal G,
 \end{cases}
\]
where $Q_{\mathbf w}$ is defined as in Lemma~\ref{lem:fixed-multiplicities} for the given $\mathbf w$
and order $r$. This gives
\[
 \begin{aligned}
 |F(1)-F(0)|
 &\le m\int_0^1
 \E_{\mathbf w\sim\pi_u^{\otimes(m-1)}}\left[
 \mathbf1_{\mathcal G}
 \left|\sum_{s\in S}\dot\pi_u(s)
       \bigl(q_{\mathbf w}(s)-P_{\mathbf w}(s)\bigr)\right|
 \right]\,du\\
 &\quad+m\int_0^1\norm{\dot\pi_u}_1
 \Pr_{\mathbf w\sim\pi_u^{\otimes(m-1)}}
       [\mathbf w\notin\mathcal G]\,du.
 \end{aligned}
\]

On $\mathcal G$, Lemma~\ref{lem:fixed-multiplicities} bounds
the residual inside the expectation above by %
\[
 \begin{aligned}
 \left|\sum_{s\in S}\dot\pi_u(s)
             \bigl(q_{\mathbf w}(s)-P_{\mathbf w}(s)\bigr)\right|
 &\le\sum_{s\in S}(\pi_0(s)+\pi_1(s))
       |q_{\mathbf w}(s)-Q_{\mathbf w}(s)|\\
 &=\sum_{b=0}^1\sum_{s\in S}\Pr[Z_b=s]\,
       |q_{\mathbf w}(s)-Q_{\mathbf w}(s)|\\
 &=\sum_{b=0}^1
       \E[|q_{\mathbf w}(Z_b)-Q_{\mathbf w}(Z_b)|]\\
 &\le\frac{4A^{r-1}}{(2r)!}
       \left(\frac{128T^2}{N}\right)^r.
 \end{aligned}
\]
where the first inequality uses the fact that $\dot\pi_u=\pi_1-\pi_0$.

Under $\pi_u^{\otimes(m-1)}$, the remaining draws are independent,
bounded by $A$, and have mean one. Thus $\E[w]=m-1\le N/2$, and
Hoeffding's inequality~\cite{Hoeffding63} gives, for every $u$,
\begin{equation}\label{eq:background-tail}
 \Pr_{\mathbf w\sim\pi_u^{\otimes(m-1)}}[\mathbf w\notin\mathcal G]
 \le\exp\left(-\frac{N}{4A^2}\right).
\end{equation}
Also, $\norm{\dot\pi_u}_1\le\norm{\pi_1}_1+\norm{\pi_0}_1 = 2$.
Substituting these bounds into the comparison above and using
$m\le N/2$ gives
\begin{equation}\label{eq:quantitative}
 \begin{aligned}
 &\left|\Pr_{f\sim\Pi_{0,N}}[\mathcal A^f=1]
       -\Pr_{f\sim\Pi_{1,N}}[\mathcal A^f=1]\right|
 =|F(0)-F(1)|\\
 &\le m\int_0^1\left(
       \frac{4A^{r-1}}{(2r)!}\left(\frac{128T^2}{N}\right)^r
       +2\exp\left(-\frac{N}{4A^2}\right)\right)\,du\\
 &\le\frac{2\times 128^r A^{r-1}}{(2r)!}\frac{T^{2r}}{N^{r-1}}
       +N\exp\left(-\frac{N}{4A^2}\right).
 \end{aligned}
\end{equation}
Using $(2r)!\ge(2r/e)^{2r}$ and $2\le2^r$, the first term is at most
\[
 \frac NA\left(\frac{64e^2AT^2}{r^2N}\right)^r.
\]
This proves~\eqref{eq:proportional-polynomial}
with $C=64e^2$ and $c=1/8$. This completes the proof.
\end{proof}

\section{Stronger bound for fixed support} %
\label{ter:sec:direct}

We prove Theorem~\ref{thm:proportional-differences} by the
same comparison lemma as the polynomial estimate (Lemma~\ref{lem:path-moments}). However, instead of using the Taylor bound, we now apply a truncated Newton expansion. We first give the
main proof using the coefficient bound of Lemma~\ref{ter:lem:average}
and the interpolation paths of Lemma~\ref{ter:lem:regularization}, which are
proved in Subsections~\ref{ter:sec:quantum} and~\ref{ter:sec:paths}, respectively.

\subsection{Proof of Theorem~\ref{thm:proportional-differences}}
\label{ter:sec:proof}

Let us fix a $T$-query quantum algorithm $\mathcal A$ on functions $f:[N]\to[N]$,
with acceptance probability $p(f)=\Pr[\mathcal A^f=1]$.
Similarly to the previous section, let 
$S=\{0\}\cup[A]$, $m=\lfloor N/2\rfloor$, 
and let $\pi_b$ be the distribution
of $Z_b$ for $b\in\{0,1\}$.
For $\mathbf z\in S^m$, let $\mathbf c(\mathbf z)$ be the frequency vector constructed in Definition~\ref{def:input-distributions}, and define
$h(\mathbf z)=\E_{f\sim\cD_{\mathbf c(\mathbf z)}}[p(f)]$.
For $\mathbf w\in S^{m-1}$, let $w=\sum_iw_i$ and
$q_{\mathbf w}(s)=h(\mathbf w,s)$ for $s\in S$.

Let us write the Newton coefficients of $q_{\mathbf w}$:
\begin{equation}\label{eq:newtion-coef}
 d_\ell(\mathbf w)=\sum_{s=0}^\ell(-1)^{\ell-s}\binom\ell s
                         q_{\mathbf w}(s)\quad(0\le\ell\le A).
\end{equation}
Let $\alpha_j=1-2^{1-j}$ for $j\ge1$ and
$\mathcal G=\{\mathbf w\in S^{m-1}:\sum_iw_i\le3N/4\}$.
The following two lemmas provide the averaged residual bound and a
path along which it is integrable. They are proved in Subsections \ref{ter:sec:quantum} and \ref{ter:sec:paths}, respectively. %

\begin{lemma}
\label{ter:lem:average}
Assume $N\ge8A$, $T\ge1$, and $T^2\le N/8$.
Let $2\le\ell\le A$ satisfy $T\le N^{\alpha_\ell/(1+\alpha_\ell)}$.
Let $\pi$ be a mean-one distribution on $S$ with $\pi(s)\ge a>0$ for every $s\in S$. There is an absolute constant $C$ such that
\[
 \begin{aligned}
 \E_{\mathbf w\sim\pi^{\otimes(m-1)}}\left[\mathbf1_{\mathcal G}(\mathbf w) \times
 \abs{d_\ell(\mathbf w)}\right]
 \le\frac{C^\ell T^{1+\alpha_\ell}}{N}(ma)^{-\alpha_\ell}, \qquad
 \Pr[\mathcal G^c]\le e^{-N/(4A^2)}.
 \end{aligned}
\]
\end{lemma}

\begin{lemma}
\label{ter:lem:regularization}
Let $2\le r\le A$ and let $\pi_0,\pi_1$ be two distinct distributions on
$\{0\}\cup[A]$ with matching moments through order $r-1$, and denote $\delta=\norm{\pi_1-\pi_0}_1$.
There is a distribution $\omega$ with these same moments such that
\[
 \omega(s)\ge32^{-A}\delta\quad(s\in\{0\}\cup[A]),
 \qquad \norm{\omega-\pi_b}_1\le\delta\quad(b=0,1).
\]
\end{lemma}

Using these lemmas, we now prove Theorem~\ref{thm:proportional-differences}.
\begin{proof}[Proof of Theorem~\ref{thm:proportional-differences}]
By Proposition~\ref{prop:normalization}, the distributions $\pi_b$ of $Z_b$
have mean one and matching moments through order $r-1$.
We know that $\delta=\norm{\pi_1-\pi_0}_1\le2$.
The conclusion is immediate if $\pi_0=\pi_1$ or $T=0$; hence let us assume $\delta>0$, and $T\ge1$.
If $r>A$, then it must be that $\pi_0=\pi_1$ since a distribution on a domain of size $A+1$ is uniquely determined by its first $A$ moments.
Hence we assume $2\le r\le A$ as well.
We choose the constants in the theorem so that $N\ge8A$ and $T^2\le N/8$.
The claim trivially follows when $T^{1+\alpha_r}>N^{\alpha_r}$.
Otherwise,
\[
 T\le N^{\alpha_r/(1+\alpha_r)}
       \le N^{\alpha_\ell/(1+\alpha_\ell)}\quad(\ell\ge r),
\]
so Lemma~\ref{ter:lem:average} applies to every remaining coefficient.

Let $\omega$ be given by Lemma~\ref{ter:lem:regularization}, so it has the same first $r-1$ moments as $\pi_0$ and $\pi_1$, and define
$\kappa=32^{-A}$. For each $b\in\{0,1\}$, we use the linear path
of probability distributions on $S$ given by
\[
 \pi_{b,u}=(1-u)\pi_b+u\omega,\qquad
 F_b(u)=\E_{\mathbf z\sim\pi_{b,u}^{\otimes m}}[h(\mathbf z)]\quad(0\le u\le1).
\]
Using the fact that the first $(r-1)$ moments of $\pi_b$ and $\omega$ match, we get that 
they also match the moments of $\pi_{b,u}$, 
which are therefore independent of $u$. Also 
\[
 \pi_{b,u}(s)\ge u \omega(s)\ge u\kappa\delta\quad(s\in S),\qquad
 \norm{\dot\pi_{b,u}}_1=\norm{-\pi_{b}+\omega}_1\le\delta.
\]
Here $F_b(0)=\Pr_{f\sim\Pi_{b,N}}[\mathcal A^f=1]$, and both paths
end at $F_b(1)=\E_{\mathbf z\sim\omega^{\otimes m}}[h(\mathbf z)]$.
We can bound the difference $|F_b(1)-F_b(0)|$ for each $b\in\{0,1\}$ using
Lemma~\ref{lem:path-moments}. We apply the lemma  with $\mathcal G=\{\mathbf w\in S^{m-1}:\sum_iw_i\le3N/4\}$ and $P_{\mathbf w}(s)=\sum_{\ell=0}^{r-1}d_\ell(\mathbf w)\binom s\ell$ of degree $r-1$ (where Newton coefficients $d_\ell(\mathbf w)$ are defined in~\eqref{eq:newtion-coef}) to upper bound $|F_b(1)-F_b(0)|$ by %
\begin{equation} \label{eq:long_UB}
m\int_0^1
 \E_{\mathbf w\sim\pi_{b,u}^{\otimes(m-1)}}
 \left[\mathbf1_{\mathcal G}
 \left|\sum_{s\in S}\dot\pi_{b,u}(s)
       \bigl(q_{\mathbf w}(s)-P_{\mathbf w}(s)\bigr)\right|\right]\,du
       +m\int_0^1\norm{\dot\pi_{b,u}}_1
 \Pr_{\mathbf w\sim\pi_{b,u}^{\otimes(m-1)}}[\mathbf w\notin\mathcal G]\,du.
\end{equation}
We now restrict our attention to bounding the two terms in the above equation.

We apply the Newton expansion in Lemma~\ref{lem:newton} to obtain
\[
 \left|\sum_{s\in S}\dot\pi_{b,u}(s)
       \bigl(q_{\mathbf w}(s)-P_{\mathbf w}(s)\bigr)\right|
 \le\delta\sum_{\ell=r}^A\binom A\ell|d_\ell(\mathbf w)|.
\]
For $u>0$, we apply Lemma~\ref{ter:lem:average} with distribution $\pi_{b,u}$ and $a=u\kappa\delta$ to get $\E\left[\mathbf1_{\mathcal G}|d_\ell(\mathbf w)|\right]
 \le\frac{C^\ell T^{1+\alpha_\ell}}{N}(ma)^{-\alpha_\ell}$ and thus 
 $$\E_{\mathbf w\sim\pi_{b,u}^{\otimes(m-1)}}
 \left[\mathbf1_{\mathcal G}
 \left|\sum_{s\in S}\dot\pi_{b,u}(s)
       \bigl(q_{\mathbf w}(s)-P_{\mathbf w}(s)\bigr)\right|\right]\ \le 
       \delta\sum_{\ell=r}^A\binom A\ell\frac{C^\ell T^{1+\alpha_\ell}}{N}(mu\kappa\delta)^{-\alpha_\ell}.$$
We absorb $\binom A\ell\kappa^{-\alpha_\ell}$ into $C^A$ by increasing
the absolute constant. The expression inside the integral of the first term
in~\eqref{eq:long_UB} is then at most
\[
 \frac{C^A}{N}\sum_{\ell=r}^A
       \frac{T^{1+\alpha_\ell}\delta^{1-\alpha_\ell}}
            {m^{\alpha_\ell}}u^{-\alpha_\ell}.
\]
The exceptional-event probability (i.e. the probability of $\mathbf{w}\notin\mathcal{G}$ in the second term) is at most $e^{-N/(4A^2)}$.
As $\alpha_\ell<1$, integration in~\eqref{eq:long_UB}
and the triangle inequality at the common endpoint give
\[
 \begin{aligned}
 \left|\Pr_{f\sim\Pi_{1,N}}[\mathcal A^f=1]
       -\Pr_{f\sim\Pi_{0,N}}[\mathcal A^f=1]\right|
 \le\sum_{b=0}^1|F_b(1)-F_b(0)|
 \le\frac{C^A m}{N}\sum_{\ell=r}^A
       \frac{T^{1+\alpha_\ell}\delta^{1-\alpha_\ell}}
            {m^{\alpha_\ell}(1-\alpha_\ell)}
       +2m\delta e^{-N/(4A^2)}.
 \end{aligned}
\]
Consider $\delta\le2$, $N/3\le m\le N/2$, and
$(1-\alpha_\ell)^{-1}=2^{\ell-1}$.
Since $T\le N$ and $\alpha_\ell\ge\alpha_r$ for $\ell\ge r$, $\frac{T^{1+\alpha_\ell}}{N^{\alpha_\ell}}
 \le\frac{T^{1+\alpha_r}}{N^{\alpha_r}}$.
Absorbing the finite sum and its exponential factors into $C^A$ yields
\[
 \left|\Pr_{f\sim\Pi_{1,N}}[\mathcal A^f=1]
       -\Pr_{f\sim\Pi_{0,N}}[\mathcal A^f=1]\right|
 \le C^A\frac{T^{1+\alpha_r}}{N^{\alpha_r}}+2N e^{-N/(4A^2)}.
\]
The left-hand side is at most one, so we may replace
$2N e^{-N/(4A^2)}$ by $N e^{-N/(8A^2)}$:
the latter is either at least one or dominates the former, since $N\ge2$. This completes the proof.
\end{proof}

\subsection{Proof of Lemma \ref{ter:lem:average}} %
\label{ter:sec:quantum}
Recall that we have fixed a $T$-query quantum algorithm $\mathcal A$ on functions $f:[N]\to[N]$, with acceptance probability $p(f)=\Pr[\mathcal A^f=1]$.

We prove the upper bound on the expectation of the Newton coefficients in Lemma~\ref{ter:lem:average} by first bounding each Newton coefficient $d_\ell(\mathbf w)$ and then averaging over $\mathbf w$. We call the case $\mathbf w\in\{0,1\}^{m-1}$ the \emph{singleton case}. We treat the singleton case separately in Lemma~\ref{ter:lem:singletons}. For the other cases, we first introduce some new notation.

\subsubsection{Newton coefficients as block differences}
\label{ter:sec:coefficients}
Assume $N\ge8A$, $2\le\ell\le A$, and fix
$\mathbf w\in S^{m-1}$ with $\sum_iw_i\le3N/4$ and some $w_i\ge2$.
Let $\mathbf c=\mathbf c(\mathbf w,\ell)$.
For notational convenience, in this subsection we identify input $f$ with its value vector $(f(1),\dots,f(N))\in[N]^N$, so coordinates of $f$ are oracle query positions.
The parts of $[N]$ with the same output label are called \emph{blocks}, their sizes correspond to the non-zero frequency entries. 

For $0\le s\le\ell$, the sampled inputs $f$ corresponding to $q_{\mathbf w}(s)$ have
the same block sizes as those of $q_{\mathbf w}(\ell)$, except
that the $\ell$-block corresponding to the last sampled multiplicity
is replaced by an $s$-block and
$\ell-s$ singletons (all singletons when $s=0$).

Let $\rho=|\{a\in[N]:c_a=\ell\}|$ be the number of blocks of size $\ell$,
and $d=|\{a\in[N]:c_a>0\}|$ the total number of blocks.
The completion is feasible since $\sum_iw_i+\ell\le7N/8$,
and $\rho\ge1$ and $d+\ell\le N$.

For $1\le j\le\ell$, starting from the fixed frequency vector $\mathbf c$, we construct a joint distribution $\nu_j$ by sampling an input $f$, a marked $j$-block $B$, and fresh replacement labels $\mathbf a$.
To sample from $\nu_j$, we first draw a uniform
partition of $[N]$ with block sizes given
by $\mathbf c$, and mark a uniform $\ell$-block.
Now let us choose a uniform $j$-subset $B$
of the marked block, split its $\ell-j$ other positions into singletons,
and give all blocks distinct uniform labels in $[N]$, finally obtaining $f:[N]\to[N]$. 
There are $d+\ell-j\le N-j$ blocks, so at least $j$ labels are unused.
Choose a uniform $j$-tuple $\mathbf a=(a_i)_{i\in B}$ of distinct
fresh labels.
We write $\E_{\nu_j}$ for expectation with respect to this distribution.

For $U\subseteq B$, let us define $f^{U,\mathbf a}$ such that
\[
 f^{U,\mathbf a}(i)=
 \begin{cases}
 a_i,&i\in U,\\
 f(i),&i\notin U.
 \end{cases}
\]
The same tuple $\mathbf a$ is used for every $U\subseteq B$,
so the inputs $f^{U,\mathbf a}$ are generated jointly.
In particular, for $j\ge2$, if $(f,B,\mathbf a)\sim\nu_j$ and
$i$ is chosen uniformly in $B$, then
$\bigl(f^{\{i\},\mathbf a},\,B\setminus\{i\},\,
       (a_k)_{k\in B\setminus\{i\}}\bigr)\sim\nu_{j-1}$,
as shown in the proof of \Cref{ter:lem:resampling}.
Together with the original triple $(f,B,\mathbf a)$, this gives a coupling between
$\nu_j$ and $\nu_{j-1}$.

The marginal distribution of $f$ under $\nu_j$ is
$\cD_{\mathbf c(\mathbf w,j)}$. In particular, it is $\cD_{\mathbf c}$
when $j=\ell$.
Thus
\[
 \E_{\nu_j}[p(f)]=q_{\mathbf w}(j).
\]

For any scalar- or vector-valued function $g$ of the input $f:[N]\to[N]$, let us now define the \emph{block difference operator} by %
\[
 D_B^{\mathbf a}g(f)=\sum_{U\subseteq B}(-1)^{|U|}g(f^{U,\mathbf a}).
\]
We henceforth suppress $\mathbf a$ from the notation for better readability.

The next lemma identifies $d_\ell(\mathbf w)$ (the Newton coefficients of $q_{\mathbf w}$) with the average block difference of the acceptance probability $\E_{\nu_\ell}[(D_Bp)(f)]$.
\begin{proposition}
 \label{lem:newton_block_difference}
For every $\ell$ and $\mathbf w$ satisfying the assumptions stated above, we have:
\begin{equation}\label{ter:eq:coefficient-identity}
 \E_{\nu_\ell}[(D_Bp)(f)]
 =d_\ell(\mathbf w).
\end{equation}
\end{proposition}
\begin{proof}
In the definition of $D_B$, taking $f^U$ for $U\subseteq B$ leaves
an $(\ell-|U|)$-block and $|U|$ singletons in place of $B$;
when $U=B$, all $\ell$ positions become singletons.
Let us take the expectation over $\nu_\ell$ and group the subsets $U\subseteq B$ by their size. There are $\binom{\ell}{|U|}$ subsets of size $|U|$, and averaging over all subsets of this size makes the split partition uniform
with these block sizes by \Cref{ter:lem:resampling}. So the resulting mean acceptance is
$q_{\mathbf w}(\ell-|U|)$. Consequently, summing over the different block sizes and denoting $s=\ell-|U|$ we obtain the following expression:
$$
 \E_{\nu_\ell}[(D_Bp)(f)]
 =\sum_{s=0}^\ell(-1)^{\ell-s}\binom\ell s q_{\mathbf w}(s)
 =d_\ell(\mathbf w).
$$%
\end{proof}

Thus, for proving Lemma~\ref{ter:lem:average}, it suffices to bound the average block difference of the acceptance
probability.

\subsubsection{Bounding the block differences}

Consider a feasible frequency vector $\mathbf c$ with $\rho\ge1$ and $d+\ell\le N$.
For the bounds below, we use notations $b_j=|\{a\in[N]:c_a=j\}|$ is the number of blocks of size $j$
for $1\le j<\ell$.
Let $n_j=1+b_j$ for $1\le j<\ell$, and $n_*=\min_{1\le j<\ell}n_j$.

The following identity of the block difference operator is going to be useful in our proofs.
For $J\subseteq B$, $D_J$ uses the labels $a_i$ with $i\in J$,
also when evaluated at $f^{\{i\}}$ for $i\in B\setminus J$.

\begin{proposition}\label{ter:lem:resampling}
Let $1\le j\le\ell$ and $(f,B,\mathbf a)\sim\nu_j$.
For a fixed $U\subseteq B$, the input $f^U$ has distinct uniform
labels on the partition obtained by splitting the positions in $U$
into singletons.

Moreover, for any scalar- or vector-valued function $g$ and every $i\in B$,
the following identity holds for every sampled triple $(f,B,\mathbf a)$, using the same
replacement labels $\mathbf a$ throughout:
\begin{equation}\label{ter:eq:factor}
 D_Bg(f)=D_{B\setminus\{i\}}g(f)
       -D_{B\setminus\{i\}}g(f^{\{i\}}).
\end{equation}
\end{proposition}
\begin{proof}
Conditional on the partition and $B$, the block labels and the
replacement labels are jointly uniform and distinct.
For a fixed $U\subseteq B$, the input $f^U$ uses a subcollection
of these labels, so its block labels are also uniform and distinct.

For the identity, let us fix a sampled triple $(f,B,\mathbf a)$ and
$i\in B$. Now we separate the subsets according to whether they contain $i$:
\[
\begin{aligned}
 D_Bg(f)
 &=\sum_{U\subseteq B\setminus\{i\}}(-1)^{|U|}
   \bigl(g(f^U)-g(f^{U\cup\{i\}})\bigr)\\
 &=D_{B\setminus\{i\}}g(f)
   -D_{B\setminus\{i\}}g(f^{\{i\}}).
\end{aligned}
\]
The last equality uses $(f^{\{i\}})^U=f^{U\cup\{i\}}$,
since all replacements use the same tuple $\mathbf a$.
\end{proof}

In the proof of the next lemma, the main fact used is that, conditional on $f$, the marked block $B$ is uniform among its
blocks of size $j$.
The first statement provides a way to go from a level $j$ quantity to a level $j-1$ quantity. Its proof is based on a double-counting bijection between level $j$ and level $j-1$. 
The second statement gives an explicit bound. %
When applying this lemma, $\ket{G(f)}$ is going to represent the function mapping input $f$ to the corresponding state of the algorithm with query access to $f$.

\begin{lemma}%
\label{ter:lem:geometry}
Let $\ket{G(f)}=\sum_i\ket{i}\otimes\ket{G_i(f)}$ be a vector-valued function of the input. Then
\begin{equation}\label{ter:eq:split}
 \E_{\nu_j} \left[\sum_{i\in B}
   \norm{D_{B\setminus\{i\}}\ket{G_i(f^{\{i\}})}}_2^2\right]
 \le\frac j{n_1}\E_{\nu_{j-1}}\left[\norm{D_B\ket{G(f)}}_2^2\right]
 \quad(2\le j\le\ell).
\end{equation}
Moreover, if $\norm{\ket{G(f)}}_2=1$ for every $f$, then
\begin{equation}\label{ter:eq:localization}
 \E_{\nu_j}\left[\sum_{i\in B}
       \norm{D_{B\setminus\{i\}}\ket{G_i(f)}}_2^2\right]
 \le\frac{4^j}{n_*}\quad(2\le j<\ell).
\end{equation}
\end{lemma}
\begin{proof}
We provide the proofs of the two statements separately.

\emph{Proof of the first statement.}
Choose $i$ uniformly in $B$ and replace $f(i)$ by $a_i$.
The modified input $f^{\{i\}}$, the marked block $B\setminus\{i\}$, and the remaining fresh labels in $\mathbf a$ correspond to the level $(j-1)$ distribution. Indeed, the old block label remains in use,
and the remaining fresh tuple is uniform among labels unused in the new input.
Conditional on this new experiment, $i$ is uniform among its
$s_j=b_1+\ell-j+1\ge n_1$ unmarked singletons: merging any one of them
back into the mark by changing its label, and then setting $a_i$ to its former label, gives a unique
preimage of the same probability.
Writing $\mathcal S$ for these singleton positions gives
\[
 \begin{aligned}
 \E_{\nu_j}\left[\sum_{i\in B}
        \norm{D_{B\setminus\{i\}}\ket{G_i(f^{\{i\}})}}_2^2\right] =\frac j{s_j}\E_{\nu_{j-1}}\left[\sum_{i\in\mathcal S}\norm{D_B\ket{G_i(f)}}_2^2\right]
 \le\frac j{n_1}\E_{\nu_{j-1}}\left[\norm{D_B\ket{G(f)}}_2^2\right].
 \end{aligned}
\]

\emph{Proof of the second statement.}
Note that $D_{B\setminus\{i\}}\ket{G_i(f)}$ is a signed sum of $2^{j-1}$ vectors $\ket{G_i(f^V)}$ (for each $V\subseteq B\setminus\{i\}$). Thus, applying Cauchy--Schwarz for each $i\in B$ gives
\[
 \sum_{i\in B}\norm{D_{B\setminus\{i\}}\ket{G_i(f)}}_2^2
 \le2^{j-1}\sum_{U\subsetneq B}\sum_{i\in B\setminus U}
                                      \norm{\ket{G_i(f^U)}}_2^2.
\]
Now let us group the terms by $s=|B\setminus U|$. Conditional on the resulting input $f^U$, the remaining marked
block $B\setminus U$ is uniform among its blocks of size $s$.
Indeed, applying on the input a permutation of positions and one of labels simultaneously preserves the
experiment; and if the pair of permutations does not change the input, it acts transitively on
its blocks of a given size.
In either case, the marked block's expected query mass is at most $1/n_*$, since
the total query mass is one and there are at least $n_s\ge n_*$ blocks of size $s$. Thus the last display has expectation at most
$2^{j-1}\sum_{s\in[j]}\binom{j}{s}/n_* = 2^{j-1}(2^j-1)/n_*\le4^j/n_*$, proving~\eqref{ter:eq:localization}.
\end{proof}

Now we turn to the main lemma of this part, that upper bounds the expected block difference of the acceptance probability. 
Our argument is an elementary form of the more involved query-gain argument
in~\cite{Belovs26}. %
\begin{lemma}
\label{ter:lem:quantum}
Under the preceding  %
assumptions, the acceptance probability
$p(f)$ of every $T$-query quantum algorithm on functions $f:[N]\to[N]$ satisfies
\begin{equation}\label{ter:eq:quantum}
 \left|\E_{\nu_\ell}[(D_Bp)(f)]\right|
 \le\frac{C^\ell T^{1+\alpha_\ell}}
             {\sqrt\rho\,n_1n_*^{\alpha_\ell-1/2}},
\end{equation}
where $C$ is an absolute constant.
\end{lemma}
\begin{proof}
By deferring measurements and uncomputing, we are going to make the accepting amplitude linear in the final state, which costs a factor 2 in queries. 
Let $A_f$ denote the whole unitary transformation, including the queries to $f$, which is applied to the initial state, $\ket{0}$. Let $P_{\mathrm{acc}}$ denote the accepting projector.
Let $V_f=A_f^\dagger(I-2P_{\mathrm{acc}})A_f$. Since $P_{\mathrm{acc}}$ is a projector, $V_f$ is unitary. Moreover
\[
 g(f):=\bra{0}V_f\ket{0}=1-2p(f).
\]
Since $V_f$ involves both $A_f$ and $A_f^\dagger$, it makes $2T$ queries to $f$, $T$ of them being reverse queries. Nonetheless, these can be rewritten as forward ones by inserting some unitary. Indeed, consider the query model $\mathcal{O}_f:\ket{i,y}\mapsto \ket{i,y+f(i) \mod N}$.
Then using the unitary $\mathrm{Neg}\ket{y}=\ket{-y\bmod N}$, we get
$\mathrm{Neg} \mathcal O_f \mathrm{Neg}\ket{i,y}=\ket{i,y-f(i) \mod N}=\mathcal O_f^\dagger\ket{i,y}$. 

So we consider $T'=2T$ forward queries in $V_f$. 
Since $D_B1=0$, we get $D_Bp=-\tfrac12D_Bg$, 
thus it suffices to bound 
$|\E_{\nu_\ell}[D_Bg(f)]|$ 
for any $T'$-query unitary circuit.
Let us write
$V_f=W_{T'}\mathcal O_fW_{T'-1}\cdots W_1\mathcal O_fW_0$. %
Absorb the final unitary $W_{T'}$ into $\bra{u}=\bra{0}W_{T'}$, and use the averaged norm
\[
 \norm{\ket{G}}_{\nu_j}
 =\left(\E_{\nu_j}\left[\norm{\ket{G}}_2^2\right]\right)^{1/2}.
\]

When there is no ambiguity, we suppress the input $f$ in the notation for states; expressions involving $D_B$ also depend on the sampled $B$ and $\mathbf a$.

Let $\ket{\psi_t}$ (or explicitly, as a vector-valued function of the input, $\ket{\psi_t(f)}$) be the state after query $t$ and $\ket{\psi'_t}$
the state before query $t+1$, so
$\ket{\psi_{t+1}}=\mathcal O_f\ket{\psi'_t}$. 
For $1\le j<\ell$, define $$R_j=\max_{0\le t\le T'}\norm{D_B\ket{\psi_t}}_{\nu_j}.$$
Since input-independent unitaries $W_t$ commute with $D_B$ and preserve these norms, we have $\|D_B\ket{\psi_t}\|_{\nu_j}= \|D_B\ket{\psi'_t}\|_{\nu_j}$. Moreover, $D_B$ applied to a constant function gives 0, thus $D_B\ket{\psi_0}=0$.
The following claim, which we prove by induction after completing the proof of the lemma,
bounds this $R_j$ as follows.
\begin{claim}\label{claim:claim}
For every $1\le j<\ell$, consider $R_j$ as defined above. Then
\[
 R_j\le\frac{4^{j+1}(T')^{\alpha_{j+1}}}
             {\sqrt{n_1}\,n_*^{\alpha_{j+1}-1/2}}.
\]
\end{claim}

It remains to bound the expected block difference at size $\ell$. We first control the contribution of each query using differences on blocks of size $\ell-1$. We then sum these contributions over the $T'$ queries and average over the uniformly chosen marked
$\ell$-block to obtain
\[
 \left|\E_{\nu_\ell}[(D_Bg)(f)]\right|
 \le \frac{2T'}{\sqrt{\rho}}\sqrt{\frac{\ell}{n_1}}\,R_{\ell-1}.
\]
Claim~\ref{claim:claim} then gives the desired bound.

\emph{Contribution of one query.} 
Let $O_i(f)$ be the oracle action on the answer register
when the queried position is $i$, so
$O_i(f)\ket{y}=\ket{y+f(i)\bmod N}$ and
$\mathcal O_f=\sum_i\ketbra{i}{i}\otimes O_i(f)$.
Let $\ket{\psi'_t(f)}=\sum_i\ket{i}\otimes\ket{\psi'_{t,i}(f)}$.
By \Cref{ter:lem:resampling}, the difference vector 
\begin{equation}\label{eq:query_diff}
 \ket{e_{B,t}}=D_B\ket{\psi_{t+1}}-\mathcal O_fD_B\ket{\psi'_t}
\end{equation}
has components
\[
 \ket{e_{B,t,i}}=
 \begin{cases}
 (O_i(f)-O_i(f^{\{i\}}))
 D_{B\setminus\{i\}}\ket{\psi'_{t,i}(f^{\{i\}})},&i\in B,\\
 0,&i\notin B.
 \end{cases}
\]
Each vector $\ket{e_{B,t,i}}$ corresponds to the component of $\ket{e_{B,t}}$ with query register in state $\ket{i}$. Thus the vectors $\ket{e_{B,t,i}}$ are orthogonal and the oracle difference has norm
at most two. Thus~\eqref{ter:eq:split} gives
\begin{equation}\label{ter:eq:error}
\begin{aligned}
 \norm{\ket{e_{B,t}}}_{\nu_j}^2
 &\le4\E_{\nu_j}\left[\sum_{i\in B}
  \norm{D_{B\setminus\{i\}}\ket{\psi'_{t,i}(f^{\{i\}})}}_2^2\right]\\
 &\le\frac{4j}{n_1}R_{j-1}^2\qquad(2\le j\le\ell).
\end{aligned}
\end{equation}

\emph{Summing over all queries.}
We now take $j=\ell$ and express $(D_Bg)(f)$ as a sum of the
query contributions, using the matrix element defining $g$.
The difference will be paired with the fixed vector $\ket{u}$,
so the computation can stay linear. This allows the uniform
choice of $B$ among $\rho$ blocks of size $\ell$ to be exploited to gain a factor $1/\sqrt{\rho}$.

Let $Q_t(f)=\mathcal O_fW_{T'-1}\cdots\mathcal O_fW_{t+1}$ be the part of the circuit after query $t+1$, with $Q_{T'-1}=I$, and let $\ket{\chi_t(f)}=Q_t(f)^\dagger\ket{u}$ be the unit vector
obtained by propagating $\ket{u}$ backwards through the circuit $Q_t(f)$.
Telescoping the query errors from \eqref{eq:query_diff} gives $D_B\ket{\psi_{T'}}=\sum_tQ_t(f)\ket{e_{B,t}} $. Therefore, recalling that $g(f)=\bra{0}V_f\ket{0}=\braket{u}{\psi_{T'}}$,
\[
 \E_{\nu_\ell}\left[(D_Bg)(f)\right]
 =\sum_{t=0}^{T'-1}\E_{\nu_\ell}\left[\langle\chi_t\mid e_{B,t}\rangle\right].
\]
Since $\ket{e_{B,t}}$ lives on query addresses in $B$, we have $\braket{\chi_t}{e_{B,t}}=\braket{P_B\chi_t}{e_{B,t}}$, where $P_B$ projects the query address onto $B$.
For a fixed input $f$ 
with its mark forgotten,
$\ket{\chi_t(f)}$ is fixed and $B$ is uniform among the $\rho$ disjoint $\ell$-blocks of $f$.
Hence $\norm{P_B\ket{\chi_t}}_{\nu_\ell}^2\le 1/\rho$, and Cauchy-Schwartz combined with \eqref{ter:eq:error} and Claim~\ref{claim:claim} finishes the proof of the lemma:
\[
 \begin{aligned}
 |\E_{\nu_\ell}[D_Bg(f)]|
 \le\sum_{t=0}^{T'-1}
       \norm{P_B\ket{\chi_t}}_{\nu_\ell}\norm{\ket{e_{B,t}}}_{\nu_\ell}
 \le\frac{2T'}{\sqrt\rho}\sqrt{\frac\ell{n_1}}\,R_{\ell-1}
 \le\frac{C^\ell(T')^{1+\alpha_\ell}}
              {\sqrt\rho\,n_1n_*^{\alpha_\ell-1/2}}.
 \end{aligned}
\]
Using $D_Bp=-\tfrac12D_Bg$ and $T'=2T$, and enlarging the absolute constant $C$, proves the lemma. %
\end{proof}

\begin{proof}[Proof of Claim~\ref{claim:claim}]
In the setting of the previous proof, for $i\in B$, let
\[
 \ket{\xi_i}=D_{B\setminus\{i\}}\ket{\psi'_{t,i}(f)},
 \qquad
 \ket{\zeta_i}=D_{B\setminus\{i\}}\ket{\psi'_{t,i}(f^{\{i\}})}.
\]
\Cref{ter:lem:resampling} gives $D_B\ket{\psi'_{t,i}(f)}=\ket{\xi_i}-\ket{\zeta_i}$ and $D_B\ket{\psi_{t+1,i}(f)}=D_BO_i(f)\ket{\psi'_{t,i}(f)}=O_i(f)\ket{\xi_i}-O_i(f^{\{i\}})\ket{\zeta_i}$.
In the last step we used that in the first term the $i$th label is always $f(i)$, and in the second it is always $a_i$. Rewrite the last expression as $O_i(f)(\ket{\xi_i}-\ket{\zeta_i}) + (O_i(f)- O_i(f^{\{i\}}))\ket{\zeta_i}$.

For any unitaries $U,V$ and vectors $\ket{\xi},\ket{\zeta}$,
\[
 \begin{aligned}
 \norm{U\ket{\xi}-V\ket{\zeta}}_2^2-\norm{\ket{\xi}-\ket{\zeta}}_2^2
 =2\operatorname{Re}\bra{\xi}(I-U^\dagger V)\ket{\zeta}
 \le4\norm{\ket{\xi}}_2\norm{\ket{\zeta}}_2.
 \end{aligned}
\]
Apply this to compare $D_B\ket{\psi_{t+1}}$ and $D_B\ket{\psi'_t}$ address by address, i.e. use $\ket{\xi}=\ket{\xi_i}$, $\ket{\zeta}=\ket{\zeta_i}$, $U=O_i(f)$ and $V=O_i(f^{\{i\}})$. Sum over $i\in B$, since outside $B$ the query leaves the difference norm unchanged. Applying Cauchy--Schwarz to the sum gives
\[
 \begin{aligned}
 \norm{D_B\ket{\psi_{t+1}}}_{\nu_j}^2-\norm{D_B\ket{\psi_t}}_{\nu_j}^2
 \le4
   \left(\E_{\nu_j}\left[\sum_{i\in B}\norm{\ket{\xi_i}}_2^2\right]\right)^{1/2}
   \left(\E_{\nu_j}\left[\sum_{i\in B}\norm{\ket{\zeta_i}}_2^2\right]\right)^{1/2}.
 \end{aligned}
\]
We prove the claim by induction on $j$, using the two bounds in
Lemma~\ref{ter:lem:geometry}. Summing the increments up to any query
time and then taking the maximum gives a bound on $R_j^2$.

For $j=1$, the first expectation is the expected query mass of $\ket{\psi'_t}$ on the marked position $i_0$ (i.e. $B=\{i_0\}$) on input $f$, which is uniform among $b_1+\ell$ singletons, so this is at most $1/(b_1+\ell)$.
The second expectation has the same interpretation, but on input $f^{\{i_0\}}$. Refreshing this label preserves the labeled input distribution, thus
both expectations in the last display are at most $1/(b_1+\ell)$.
Each increment is therefore at most $4/(b_1+\ell)\le4/n_1$. Since $D_B\ket{\psi_0}=0$, summing over $T'$ queries gives $R_1^2\le4T'/n_1$.

For $2\le j<\ell$, apply~\eqref{ter:eq:localization} to the first expectation to obtain upper bound $4^j/n_*$.
Then applying~\eqref{ter:eq:split} to the second expectation gives upper bound $\frac{j}{n_1}R_{j-1}^2$. Thus, the increment is at most $4\frac{2^j}{\sqrt{n_*}}\frac{\sqrt{j}}{\sqrt{n_1}}R_{j-1}$.
The initial differences vanish, so summing over at most $T'$ queries yields
\[
 R_j^2\le\frac{4\sqrt j\,2^jT'}{\sqrt{n_1n_*}}R_{j-1}
          \quad(2\le j<\ell).
\]
Use the induction hypothesis together with $\alpha_2=1/2$, $\alpha_{j+1}=(1+\alpha_j)/2$ and $\sqrt j\,2^j\le4^{j+1}$, to obtain
\[
 R_j\le\frac{4^{j+1}(T')^{\alpha_{j+1}}}
             {\sqrt{n_1}\,n_*^{\alpha_{j+1}-1/2}}.
\]
\end{proof}

\subsubsection{Newton coefficient bounds}
Now we return to Newton coefficients, the original goal of \Cref{ter:lem:average}.
As mentioned earlier, we first consider the singleton case, i.e., $\mathbf w\in\{0,1\}^{m-1}$.
\begin{lemma}
\label{ter:lem:singletons}
Suppose $N\ge8A$, $T\ge1$, and $T^2\le N/8$.
Let $2\le\ell\le A$ satisfy $T\le N^{\alpha_\ell/(1+\alpha_\ell)}$.
For every $\mathbf w\in\{0,1\}^{m-1}$,
\[
 |d_\ell(\mathbf w)|\le C^\ell\frac{T^{1+\alpha_\ell}}{N},
\]
where $C$ is an absolute constant.
\end{lemma}
\begin{proof}
For integer $0\le s\le N$, let $q(s)$ be mean acceptance for one
$s$-block and $N-s$ singletons, uniformly positioned and labeled.
This function agrees with $q_{\mathbf w}$ on $\{0\}\cup[A]$, and the cases $s=0,1$ both give the all-singleton partition.
Its power sums are $N-s+s^j$ for $j\ge1$, so
Lemma~\ref{lem:symmetrization} extends $q$ to a polynomial of degree
at most $2T$, bounded in $[0,1]$ at the integers $0,\ldots,N$.

Since $(2T)^2\le N/2$, we apply Lemma~\ref{lem:newton} with $H=N$,
remainder order $\ell$, and $z=\ell$.
The remainder at this point is exactly $d_\ell(\mathbf w)$, giving
\[
 |d_\ell(\mathbf w)|
 \le\frac{2\ell^\ell}{(2\ell)!}
           \left(\frac{16T^2}{N}\right)^\ell
 \le C^\ell\left(\frac{T^2}{N}\right)^\ell
 \le C^\ell\frac{T^{1+\alpha_\ell}}{N}.
\]
The last inequality follows from $T\le N^{\alpha_\ell/(1+\alpha_\ell)}$
and
\[
 (\ell-1)(1+\alpha_\ell)-\alpha_\ell(2\ell-1-\alpha_\ell)
 =(1-\alpha_\ell)(\ell-1-\alpha_\ell)\ge0.
\]
\end{proof}

Now we continue with general $\mathbf{w}$, using the block difference bounds.
We assume $N\ge8A$, $T\ge1$, and $T^2\le N/8$, and fix
$2\le\ell\le A$ with $T\le N^{\alpha_\ell/(1+\alpha_\ell)}$.

\begin{lemma}
\label{ter:lem:fixed}
Let $\mathbf w\in S^{m-1}$ satisfy $\sum_iw_i\le3N/4$.
First, let us complete $(\mathbf w,\ell)$ with singletons, let $\rho$ count its
$\ell$-blocks, and let $b_j$ count its $j$-blocks for $1\le j<\ell$.
Let $n_j=1+b_j$ and $n_*=\min_{1\le j<\ell}n_j$. Then
\begin{equation}\label{ter:eq:pointwise}
 |d_\ell(\mathbf w)|
 \le\frac{C^\ell T^{1+\alpha_\ell}}
              {\sqrt\rho\,n_1n_*^{\alpha_\ell-1/2}},
\end{equation}
where $C$ is an absolute constant.
\end{lemma}
\begin{proof}
The completion is feasible since $\sum_iw_i+\ell\le7N/8$.
If $w_i\le1$ for all $i$, Lemma~\ref{ter:lem:singletons} gives
$|d_\ell(\mathbf w)|\le C^\ell T^{1+\alpha_\ell}/N$, which proves \eqref{ter:eq:pointwise} with $\rho=1$, $n_1=N-\ell+1\le N$, and
$n_*^{\alpha_\ell-1/2}=1$: the exponent is zero for $\ell=2$,
and $n_*=1$ for $\ell\ge3$. 

Otherwise, the completed partition has at most $N-\ell$ blocks, 
thus Lemma~\ref{ter:lem:quantum} applies.
By \Cref{lem:newton_block_difference}, the expected block difference is $d_\ell(\mathbf w)$.
Using the bound of Lemma \ref{ter:lem:quantum} finishes the proof.
\end{proof}

Now we have to average over $\mathbf{w}$ to finish the proof.
\begin{proof}[Proof of Lemma~\ref{ter:lem:average}]
The independent draws have mean one and is in $S$, so
$\E [\sum_iw_i]=m-1\le N/2$.
Hoeffding's inequality~\cite{Hoeffding63} gives the stated bound on
$\Pr[\mathcal G^c]$.

Let $Y_j$ count the draws equal to $j$. Under the unconditioned product
distribution, $Y_j\sim\operatorname{Bin}(m-1,\pi(j))$.
On $\mathcal G$, at least $N-\sum_iw_i-\ell\ge N/8$ completion
singletons are present, and the counts in Lemma~\ref{ter:lem:fixed} satisfy
\[
 \rho=1+Y_\ell,\qquad n_1\ge N/8,\qquad
 n_j\ge1+Y_j\quad(1\le j<\ell).
\]
For the following bound, we first apply Lemma \ref{ter:lem:fixed} and we drop the event indicator.
We also use the following facts: since $\alpha_\ell-1/2\ge0$, on
$\mathcal G$ we have $n_*^{-\alpha_\ell+1/2}\le\sum_{j=1}^{\ell-1}(1+Y_j)^{-\alpha_\ell+1/2}$; and $\rho^{-1/2}\le(1+Y_\ell)^{-1/2}$. Thus
\[
 \begin{aligned}
 \E\left[\mathbf1_{\mathcal G}|d_\ell(\mathbf w)|\right]
 \le\frac{C^\ell T^{1+\alpha_\ell}}{N}
       \sum_{j=1}^{\ell-1}
       \E\left[(1+Y_\ell)^{-1/2}(1+Y_j)^{-\alpha_\ell+1/2}\right] \le\frac{C^\ell T^{1+\alpha_\ell}}{N}(ma)^{-\alpha_\ell}
 \end{aligned}
\]
In the last step we used Lemma~\ref{lem:inverse-binomial} (since $\alpha_\ell<1$) with the fact that $a\le\pi(s)$,
and absorbed factor $\ell-1$ into $C^\ell$.
\end{proof}

\subsection{Proof of Lemma~\ref{ter:lem:regularization}}
\label{ter:sec:paths}

We prove Lemma~\ref{ter:lem:regularization} by adding a small uniform
component and correcting its low moments on $r$ suitable support points.

\begin{proof}[Proof of Lemma~\ref{ter:lem:regularization}]
Let $v=\pi_1-\pi_0$, $\bar\pi=(\pi_0+\pi_1)/2$, and $\Lambda=4^{A-1}$.
Let $I'$ index the $r-1$ largest coordinates of $|v|$.
For its Lagrange basis and $i\in I'$, moment cancellation and
Lemma~\ref{lem:grid-interpolation} give
\[
 v(i)=-\sum_{j\notin I'}v(j)L_i(j),\qquad
 \sum_{i\in I'}|v(i)|\le \Lambda\sum_{j\notin I'}|v(j)|.
\]
Thus the $r$th largest coordinate of $|v|$ is at least
$\delta/((A+1)(1+\Lambda))$. On the set $I$ of the $r$ largest coordinates,
\[
 \bar\pi(i)\ge |v(i)|/2\ge \eta:=\frac{\delta}{2(A+1)(1+\Lambda)}.
\]
Let $U$ be uniform on $\{0\}\cup[A]$, let $L_i$ now denote the
Lagrange basis on $I$, and let $e_i$ be unit mass at $i$. Let us set
\[
 \tau=\eta/\Lambda,\qquad
 \omega=\bar\pi+\tau\left(U-\sum_{i\in I}(\E_U[L_i])e_i\right).
\]
For every polynomial $P$ of degree below $r$, interpolation gives
$\E_U[P]=\sum_{i\in I}P(i)\E_U[L_i]$.
Thus the correction has zero moments through order $r-1$, including
zero total mass.
Lemma~\ref{lem:grid-interpolation} gives $\sum_{i\in I}|\E_U[L_i]|\le \Lambda$.
Since $\tau \Lambda=\eta$, both on $I$ and outside it, we have
\[
 \omega(s)\ge\frac\tau{A+1}
 =\frac{\delta}{2(A+1)^2\Lambda(1+\Lambda)}\ge32^{-A}\delta.
\]
The last inequality uses the fact that
$2(A+1)^2\Lambda(1+\Lambda)\le4(A+1)^2 16^{A-1}\le32^A$ for $A\ge2$.
Hence $\omega$ is a probability distribution with the required moments. Finally,
\[
 \norm{\omega-\pi_b}_1
 \le\norm{\bar\pi-\pi_b}_1+\tau(1+\Lambda)
 \le\delta/2+2\eta\le\delta.
\]
\end{proof}

\section{Applications}\label{sec:applications}

In this section, we illustrate the moment matching method on various applications such as $k$-distinctness testing, testing some graph properties,
image-size testing, and junta testing. For all these problems, classical frequency-moment constructions supply the hard inputs and Theorems~\ref{thm:proportional-polynomial} and~\ref{thm:proportional-differences}
give quantum lower bounds, which we transfer to graph testing problems and junta testing
by reductions. Table~\ref{tab:applications-summary} compares the
resulting bounds with previous randomized and quantum query bounds
and identifies the argument used. 
The list of applications is not exhaustive, the purpose of this section is only to provide some interesting examples where our main theorems can be applied. 
Throughout, $k$ and testing proximity parameter $\eps$ are considered to be constants, and we use shorthand
$a_k=(2^{k-1}-1)/(2^k-1)$. Before presenting the application results, we discuss some preliminaries.

\subsection{Preliminaries for applications}\label{sec:prelimapp}

We denote $[R]_0=\{0,1,\ldots,R\}$.
In the query complexity model, we consider inputs $x\in \Sigma^I$ over a finite alphabet $\Sigma$ and indexed by a set $I$. When the input is a string $s=(s_1,\dots,s_N)$ of positive integers $\le R$, then $I=[N]$ and $\Sigma=[R]$.

In the \emph{bounded-degree graph model}, we query the adjacency lists of the vertices of the input graph. For a graph with vertex set $V$ and maximum degree bound $d$, we have $I=V\times [d]$ and $\Sigma=V\cup\{\bot\}$. Here we assume that each vertex has a fixed order of its neighbors, and a query $(v,i)$ returns the $i$-th neighbor of $v$ if it exists, and a special character $\bot$ otherwise.
For directed graphs, one can distinguish two variants of this model.
In the \emph{unidirectional model}, the algorithm only has access to the list of out-neighbors of the vertices. On the other hand, in the \emph{bidirectional model}, there are two lists: one for the out-neighbors and one for the in-neighbors, so the algorithm has to specify if it would like to learn the $i$-th in- or out-neighbor of a vertex $v$. This can be achieved, for example, by the query containing one additional bit.

In quantum query complexity, the algorithm can use a unitary query operator, and for input $x\in\Sigma^I$ one way to define it as follows: $\mathcal O_x\ket{i}\ket{z}= \ket{i}\ket{z + x_i \ \text{mod} \ |\Sigma|}$, for $z\in \Sigma$ and $i\in I$.

A \emph{property} is a predicate $\mathcal{P}:D_{\mathcal P}\to\{0,1\}$ over domain $D_{\mathcal P}\subseteq \Sigma^I$. We say that an input $x$ satisfies $\mathcal P$ if $x\in \mathcal P$.
The distance of two inputs $x,y\in \Sigma^I$ is defined as
	$$\dist(x,y)=\frac{|\{i\in I: x_i\neq y_i\}|}{|I|},$$
	We say that $x$ is \emph{$\eps$-far} from a property $\mathcal P$ if $\dist(x,y)>\eps$ for all $y\in \mathcal P$.

In the case of graphs, 
a \emph{graph property} is defined as a set of graphs that is closed under graph isomorphism.
In the case of bounded-degree graphs with maximum degree bound $d$, the distance of two graphs is the number of edges that one needs to add or delete in order to transform one to the other (up to isomorphism) divided by $|V|d$. 

In \emph{property testing}, an algorithm, also called a $\varepsilon$-\emph{tester}, accepts inputs in $\mathcal P$ and rejects those $\varepsilon$-far from $\mathcal P$, each with probability at least $2/3$. The tester can either output accept or reject when the above two conditions do not hold.

The problem of $k$-source-subgraph-freeness testing, and its special case, $k$-star-freeness testing, were introduced in \cite{HS12} and their quantum versions were studied in \cite{AMSS25}. The former is a general directed subgraph-freeness problem where the forbidden subgraph is restricted to having $k$ ``source components'', where
a source component is a set of vertices $S$ with no edge going from vertices outside $S$ to $S$.
An (incoming) $k$-star consists of a center node and $k$ outer nodes all having an edge to the center.%

We also present directed planarity and outerplanarity testing lower bounds. 
Here, planarity and outerplanarity refer to the underlying undirected graph obtained by ignoring arc directions and replacing each antiparallel edge pair by a single edge, but the oracle gives only outgoing neighbors in the unidirectional model defined above. 
Recall that an undirected graph $G$ is said to be \emph{planar} if it can be drawn in the
plane such that edges intersect only at common endpoints. Similarly, $G$ is said to be \emph{outerplanar} if it can be drawn in a planar way with every vertex on the boundary of the outer (unbounded) face.

\subsection{Testing distinctness and graph properties}

We apply \Cref{thm:proportional-differences} to the classical hard instances considered by 
Peng and Wang~\cite{peng2023optimal}, then transfer the distinctness bound to graph
testing through a local encoding.

\subsubsection{The classical frequencies give a quantum lower bound}

The classical random variables $X_0,X_1$ on $[k]$ from
\cite[Section~2.2, Lemma~11]{peng2023optimal} have $k-1$ proportional
moments. We take the classical hard distribution from \cite{peng2023optimal} and apply Theorem~\ref{thm:proportional-differences};
it remains to check that its input distributions satisfy the testing
promises.

\begin{corollary}[Testing $k$-distinctness]\label{cor:k-distinctness}
For any $k\ge2$, let us fix $\varepsilon_k=1/(k2^k)$. Also let $M\ge N$.
For every $0<\varepsilon\le\varepsilon_k$, testing
\[
 \mathcal E_{k,N,M}
 =\{f:[N]\to[M]: |f^{-1}(a)|<k\text{ for every }a\in[M]\}
\]
has quantum query complexity $\Omega(N^{a_k})$, even when every multiplicity
is at most $k$, where $a_k=(2^{k-1}-1)/(2^k-1)$. %
\end{corollary}

\begin{proof}
Let us consider $X_0,X_1$ from~\cite[Section~2.2]{peng2023optimal}. Besides their
$k-1$ proportional moments, they satisfy
\[
 \Pr[X_0=k]=0,\qquad
 \frac{\Pr[X_1=k]}{\E\,[X_1]}
 =\frac1{k2^{k-2}}=4\varepsilon_k.
\]
For $M=N$, let us use the input distribution $\Pi_{b,N}$ of Definition~\ref{def:input-distributions}. Under $\Pi_{0,N}$, every input belongs
to $\mathcal E_{k,N,N}$. In Definition~\ref{def:input-distributions},
each sampled multiplicity equals $k$ with probability
$\Pr[X_b=k]/\E\,[X_b]$. Using a concentration argument, we can say that with probability $1-e^{-\Omega(N)}$, no overflow event happens and $\Pi_{1,N}$ contains at least $\varepsilon_kN$ values of multiplicity $k$. 
Each such value requires at least
one change, so with probability $1-o(1)$, such inputs are $\eps_k$-far from $\mathcal E_{k,N,N}$, and thus a tester has acceptance gap $1/3-o(1)$.

For $T>c\sqrt N$, the lower bound is immediate. Otherwise,
Theorem~\ref{thm:proportional-differences}, with $A=r=k$, gives us
\[
 \frac13-o(1)
 \le C^k\frac{T^{2-2^{1-k}}}{N^{1-2^{1-k}}}
       +Ne^{-cN/k^2}.
\]
Thus $T=\Omega(N^{a_k})$. Embedding the outputs into $[M]$ preserves
the distance bound and costs only constantly many queries. This completes the proof.
\end{proof}

\paragraph{Previous quantum bounds}
The bound $\Omega(N^{1/2-1/(2k)}/\log^2N)$ in
\cite[Theorem~4.1]{AMSS25} comes from
dual polynomials, and
\cite[Theorem~7.2]{PW26} obtained the same bound (with a worse logarithmic factor) even for bounded multiplicities.
Corollary~\ref{cor:k-distinctness} improves this
exponent for $k\ge3$, and
our new bound matches the upper
bound from~\cite[Sections~1.3 and~1.5.1]{AMSS25}.
The optimal lower bound of $\Omega(N^{1/3})$ for $k=2$ was known from \cite{aaronson2004quantum} (since a two-to-one function is $(1/2)$-far from injectivity), our result proves optimal bounds for $k \geq 3$. 
It is interesting to note that the problem of \emph{finding} $k$-collisions in random functions has the same query complexity as the property testing variant \cite{zhandry_collision}.

\subsubsection{Graph consequences}\label{s2:sec:graph-model}
We describe the standard local reduction that maps a $k$-collision to a $k$-star; the other consequences (planarity and outerplanarity) use
the same construction. Recall that for planarity and outerplanarity testing, while the graph is directed and accessed in the unidirectional model, the property to test is whether the underlying undirected graphs are planar and outerplanar, respectively.

Let $K_{a,b}$ denote the complete bipartite graph with the two parts having size $a$ and $b$.
In its directed variant, every arc is directed from the first part to the second.
From a distinctness testing instance $f:[N']\to[N']$, 
we form a directed graph $G_q(f)$ of size $N=N'(q+1)$ with vertices $u_i$ and $v_{a,j}$,
for $i,a\in[N']$, $j\in[q]$ for some constant parameter $q$, and arcs
\begin{equation}\label{s2:eq:graph-construction}
 u_i\longrightarrow v_{f(i),j}.
\end{equation}
By this construction, each value of multiplicity $c_a$ gives a directed $K_{c_a,q}$ component, leading to a graph of maximum outdegree $q$.
Moreover, by this reduction, $k$-distinctness testing turns into different subgraph-freeness testing problems:

\begin{enumerate}
\setlength{\itemsep}{0pt}
\item For $q=1$ and any constant $k$, the reduction gives the $k$-star-freeness testing problem, which is a special case of $k$-source-subgraph-freeness testing. 

\item For $q=3$ and $k=3$, the reduction gives a planar graph in the $3$-collision-free case and a far from planar one (containing many disjoint $K_{3,3}$ subgraphs) in the far from $3$-collision-free case. 

\item For $q=2$ and $k=3$, the reduction gives an outerplanar graph in the $3$-collision-free case and a far from outerplanar one (containing many disjoint $K_{3,2}$ subgraphs) in the far from $3$-collision-free case. %
\end{enumerate}

Thus, we obtain the following corollary.

\begin{corollary}[Graph testing problems]\label{s2:thm:directedgraph}
In the unidirectional bounded-degree model, the following quantum query lower bounds hold for $N$-vertex input graphs:
\begin{enumerate}
\setlength{\itemsep}{0pt}

\item For any constant $k$ and 
$0<\varepsilon\le\varepsilon_k/4$, testing 
$k$-source-subgraph-freeness
has quantum query complexity $\Omega(N^{a_k})$. A special case is  $k$-star-freeness.

\item For $0<\varepsilon\le1/576$, testing planarity with maximum outdegree $3$ has quantum query complexity $\Omega(N^{3/7})$.

\item For $0<\varepsilon\le1/288$, testing outerplanarity with
maximum outdegree $2$ has quantum query complexity $\Omega(N^{3/7})$.
\end{enumerate}

The bounds hold already on bipartite directed acyclic graphs whose
indegree and undirected degree are at most $k$ in the first case,
and at most $3$ in the last two cases. 
\end{corollary}

\begin{proof}
Under the promise that multiplicities are at most $k$, a function that is $\varepsilon_k$-far from $k$-distinctness of input length $N'$ has at least $\varepsilon_kN'$ values of multiplicity $k$: changing one occurrence
of each such value to an unused label repairs the function. We form $G_q(f)$ similar to \cref{s2:eq:graph-construction}.
Thus $G_q(f)$ has that many disjoint forbidden components and each component requires an original arc to be deleted. With
$N'=\lfloor N/(q+1)\rfloor$ and adding isolated vertices if required, the distance
is at least $\varepsilon_kN'/(qN)\ge\varepsilon_k/[2q(q+1)]$. %
Correctness follows from the underlying component graphs in the following table. %

\[
\begin{array}{c|c|c}
\text{property}&\text{allowed components}&\text{forbidden component}\\ \hline
k\text{-star-freeness}&K_{j,1},\ j<k&K_{k,1}\\
\text{planarity}&K_{1,3},K_{2,3}&K_{3,3}\\
\text{outerplanarity}&K_{1,2},K_{2,2}&K_{3,2}
\end{array}
\]

We finally apply Corollary~\ref{cor:k-distinctness}, with the same $k$ for the first row, and $k=3$ for the last two rows.
\end{proof}

\paragraph{Previous quantum bounds}
For $k$-star-freeness and $k$-source-subgraph-freeness testing, the lower bound in~\cite{AMSS25} transfers its $k$-distinctness testing bound of $\Omega(N^{1/2-1/(2k)}/\log^2N)$ through the described local graph encoding (for $q=1$). The same lower bound was extended to the bounded indegree case by \cite{PW26}. For planarity and outerplanarity testing, using the lower bound of $\tilde\Omega(N^{1/2-1/(2k)})$, the same reduction gives a bound of $\tilde\Omega(N^{1/3})$ and, to the best of the authors' knowledge, this is the current best lower bound.

\subsection{Image-size and junta testing}

We use $r = \Theta(\log N)$ proportional moments on support $A=O(r^2)$.
The polynomial estimate gives a square-root bound for image size
testing, which can be transferred to junta testing by a well-known addressing reduction.

\subsubsection{Classical moment construction}\label{sec:classicalmomentconstruction}

Let us first revisit the classical moment construction for this problem. In fact, \cite[Theorem~4.5]{RRSS09} presented a similar construction for the distinct elements problem, which is related to image size testing. Their construction considers $A=O(\exp(r))$. However, for our applications, we need $A=O(r^2)$. We will prove the next lemma by using
polynomial approximation and linear-programming duality, following the approach of~\cite[Appendix~A]{wu2019chebyshev}. The proof is deferred to Appendix~\ref{proof:5.4}.

\begin{restatable}{lemma}{momentconstructionlemma}
\label{lem:quadratic-support}
For fixed $0<\eta<1/2$ and every $r\ge2$, there exist random variables 
$X_0,X_1$ on $[A]$, with $A\le Cr^2$, such that
\[
 \frac{\E\,[X_0^j]}{\E\,[X_0]}
 =\frac{\E\,[X_1^j]}{\E\,[X_1]}\quad(1\le j<r),\qquad
 \frac1{\E\,[X_0]}\ge1-\eta,\quad
 \frac1{\E\,[X_1]}\le\eta.
\]
Here $C$ depends only on $\eta$.
\end{restatable}

\begin{remark}
The quadratic bound on the largest multiplicity also improves the lower
bound obtained from the classical argument of \cite{RRSS09}. Their construction
\cite[Theorem~4.5]{RRSS09} uses $A=O(\exp(r))$ and their lower bound is $\Omega(N^{1-o(1)})$~\cite[Corollary~2.2]{RRSS09}.
Replacing that construction by Lemma~\ref{lem:quadratic-support}, while retaining their other arguments, gives a lower bound
$\Omega\left(N\left(\frac{\log\log N}{\log N}\right)^4\right)$.
In particular, this gives an improved lower bound on the distinct elements and support-size estimation problems, although the optimal lower bound is $\Omega(N/\log N)$.  
\end{remark}

\subsubsection{Image-size testing}
We consider inputs $f:[N]\to[N]$. 
The image-size testing problem considered in \cite{ABRW16}, corresponds to testing the property of having image size at most $\ell$ where $\ell< N$ is any parameter.
Then, \cite{BKT20} considered a special case of distinguishing image size at most $\gamma N$ for $0<\gamma<1$ from image size exactly $N$.
In this work, we consider a different special case of the version in \cite{ABRW16}, where we 
distinguish image size at most $\lfloor3N/4\rfloor$ from at least $\lfloor7N/8\rfloor$.
This hard instance provides the same lower bound for the more general variant of image-size testing.

We apply Theorem~\ref{thm:proportional-polynomial} with the random variables provided by Lemma~\ref{lem:quadratic-support}. Singleton completion preserves the gap in image sizes.

\begin{corollary}[Image-size testing]\label{cor:image-size}
Let $\mathcal P_N=\{f:[N]\to[N]:|f([N])|\le\lfloor3N/4\rfloor\}$.
Testing $\mathcal P_N$ at proximity $1/8$ has quantum query complexity
$\Omega(\sqrt N)$. 
\end{corollary}

\begin{proof}
Let us take $r=\lceil\log N\rceil$, $\eta=1/10$, and the variables $X_0,X_1$
of Lemma~\ref{lem:quadratic-support}. Now we denote $\mu_b=\E\,[X_b]$ and
$m=\lfloor N/2\rfloor$. Following Definition~\ref{def:input-distributions},
each sampled multiplicity has mean $1$ and is nonzero with probability
$1/\mu_b$. Singleton completion therefore gives us
\[
 \E_{f\sim\Pi_{b,N}}\,[|f([N])|]
 =N-m(1-1/\mu_b)+o(1),
\]
where the correction term corresponds to the overflow event, which happens with probability $o(1)$.
Using the properties of $X_0$ and $X_1$, for $b=0$ we get expected image size at least $19N/20+o(1)$, and for $b=1$, it is at most $11N/20+o(1)$. Concentration of the sampled multiplicities, together with the overflow
bound, gives $|f([N])|\ge\lfloor7N/8\rfloor$ with very high probability for the input coming from $X_0$ and $|f([N])|\le\lfloor3N/4\rfloor$ for the one coming from $X_1$. 
Removing an output value requires changing at least one position, thus $f_1\in\mathcal{P}_N$ and $f_0$ is $1/8$-far, so a
tester must have acceptance gap at least $1/3-o(1)$.

Applying Theorem~\ref{thm:proportional-polynomial} directly to $X_0,X_1$ using the fact that $A=O(r^2)$, and combining the resulting acceptance gap upper bound with the previous lower bound, we obtain $T\in\Omega(\sqrt{N})$.

\end{proof}

\paragraph{Previous quantum bounds}
Ambainis, Belovs, Regev, and de Wolf \cite{ABRW16} proved a lower bound of $\Omega(N^{1/3})$ on the general image-size testing problem using that the collision problem (distinguishing one-to-one and two-to-one functions) is a special case.
Bun, Kothari and Thaler~\cite[Theorem~6.2 and Section~6.1.2]{BKT20} improved the lower bound to $\Omega(\sqrt N/\log^2 N)$ using their special case of distinguishing image size $N$ from at most $\gamma N$. 
Our Corollary~\ref{cor:image-size} removes the $\log^2N$ term, via the special case of distinguishing image size at most $\lfloor3N/4\rfloor$ and greater than $\lfloor7N/8\rfloor$. 
Our lower bound matches the quantum upper bound from \cite[Theorem~2]{montanaro2015quantum}.

\begin{remark} 
\Cref{thm:proportional-differences} improves the lower bound from \Cref{thm:proportional-polynomial} when $A$ and $r$ are fixed, but its exponential dependence on $A$ can make it ineffective when the multiplicities grow with the input size. Image-size testing provides a concrete example. In the proof of Corollary~\ref{cor:image-size}, we take $r=\lceil\log N\rceil, A=\Theta(r^2)$. Applying \Cref{thm:proportional-differences} would yield a suboptimal lower bound of $\Omega(N^{1/2 - o(1)})$. By the reduction, the same holds for junta testing as well.
\end{remark}

\subsubsection{Junta testing by reduction}

A Boolean $K$-junta depends on at most $K$ input coordinates.
We apply the addressing reduction of~\cite[Lemma~6.3]{ABRW16} (already present in \cite{atici2007quantum})
to Corollary~\ref{cor:image-size}. 
Since this reduction is standard in the literature (see e.g., \cite[Section 6.2]{BKT20}), we only state the result. %

\begin{corollary}[Junta testing]\label{s3:thm:junta}
Testing Boolean $K$-juntas at proximity $1/3$ has quantum query
complexity $\Omega(\sqrt K)$. 
\end{corollary}

\paragraph{Previous quantum bounds}
\cite{ABRW16} proved a lower bound of $\Omega(K^{1/3})$, which was later improved by \cite{BKT20} to $\Omega(\sqrt K/\log^2 K)$. Both results used a reduction from image size testing.

\subsection*{Acknowledgements}
 
CK's and FM's research was supported in part by a bilateral grant between
the Université Paris Cité and the National University of Singapore
(2026-01-RESEARCH/UPCité-NUS and ANR-18-IDEX-0001),
the French PEPR integrated projects EPIQ (ANR-22-PETQ-0007)
and HQI (ANR-22-PNCQ-0002), and the ERC Advanced Grant PARQ.

SS's research was supported by the NRF Investigatorship award (NRF-NRFI10-2024-0006) and CQT Young Researcher Career Development Grant (25-YRCDG-SS). 

DS's research was supported by the German Federal Ministry of Research, Technology and Space (QuSol, 13N17173), by the Deutsche Forschungsgemeinschaft under Germany’s Excellence Strategy – EXC-2111 – 390814868, and by the Munich Quantum Valley.

\paragraph{AI statement.} We used GPT-6 Astra, GPT-6.1 Sol and Claude Opus 5.5 to explore proof strategies, refine mathematical arguments, search for related literature, and improve the presentation. 
The authors subsequently verified, simplified,
and organized the argument, and rewrote its exposition. Important concepts and ideas were manually factored out and written down in an intuitive, understandable way. The authors
take full responsibility for the correctness and content of the paper. 

\addcontentsline{toc}{section}{References}
\bibliographystyle{alpha}
\bibliography{biblio}

\appendix

\section{Deferred proofs on polynomials} %
\label{appendix:proof of aux results}

We provide the missing proofs of some lemmas from \Cref{sec:approx-moments-bounds}.
We restate these lemmas for ease of reading.

\taylorlemma*
\begin{proof} %
Denote $M= \norm{q}_{[0,H]}=\max_{z\in[0,H]}|q(z)|$.
Markov's inequality gives $$\norm{q'}_{[0,H]}\le \frac{2d^2}{H}M,$$ where $q'$ denotes the first order derivative of $q$.
A point attaining $M$ lies within $1/2$ of an integer, so
\[
 M\le1+\tfrac12\norm{q'}_{[0,H]}
   \le1+\frac{d^2}{H}M\le1+\frac M2,
 \qquad\text{hence }M\le2.
\]
If $r>d$, the Taylor remainder is zero. Otherwise, applying the higher-order
Markov inequality and rescaling from $[-1,1]$ to $[0,H]$, gives
\[
 \frac{\norm{q^{(r)}}_{[0,H]}}{r!}
 \le\frac{M}{r!}\left(\frac2H\right)^r
             \frac{d^{2r}}{(2r-1)!!}
 \le\frac{2}{(2r)!}\left(\frac{4d^2}{H}\right)^r,
\]
where $q^{(r)}$ denotes the $r$-th order derivative and for the last inequality we use $r!(2r-1)!!=(2r)!/2^r$~\cite{DeVoreLorentz93}. 
Applying Taylor's theorem we get~\eqref{eq:taylor}.
\end{proof}

\newtonlemma*
\begin{proof} %
First we observe that
$$q(z)=\sum_{\ell=0}^H d_\ell\binom z\ell=\sum_{\ell=0}^z d_\ell\binom z\ell.$$
Indeed, for $0\le t\le z$, substituting the definition of $d_\ell$ gives
the coefficient $\binom{z}{t}(1-1)^{z-t}$ for $q(t)$.
This is one for $t=z$ and zero otherwise, proving the expansion
and the remainder identity.

For the bound, extend $q$ to the stated polynomial.
The polynomial $Q_r$ interpolates $q$ at $0,\ldots,r-1$.
If $r>d$, both \(q\) and \(Q_r\) have degree at most \(r-1\). Since they agree at \(r\) points, they must be identical polynomials. Otherwise, the interpolation remainder
formula and the derivative estimate in the proof of
Lemma~\ref{lem:taylor} give
\[
 \begin{aligned}
 |q(z)-Q_r(z)|
 &\le\frac{\norm{q^{(r)}}_{[0,H]}}{r!}
               \left|\prod_{j=0}^{r-1}(z-j)\right|\\
 &\le\frac{2}{(2r)!}\left(\frac{4d^2}{H}\right)^r z^r.
 \end{aligned}
\]
At integer $z<r$, the product is zero; at $z\ge r$, it is at most $z^r$.
\end{proof}

\gridinterpolationlemma* 
\begin{proof} %
The first identity is the standard Lagrange interpolation
formula~\cite[\S 3.3(i)]{DLMF}.
At a node, $\sum_i|L_i(t)|=1$. Otherwise, the $h$-th largest
distance from $t$ to a grid point is at most $A-h+1$, so the
numerator of each $L_i(t)$ is at most $\falling{A}{d}$.
Integer spacing bounds its denominator below by $i!(d-i)!$.
Hence
\[
 \sum_{i=0}^d|L_i(t)|
 \le\sum_{i=0}^d\frac{\falling{A}{d}}{i!(d-i)!}
 =2^d\binom Ad\le4^{A-1}.
\]
To prove the last inequality, observe that $2^d\binom Ad$ is the
coefficient of $x^d$ in the binomial expansion
\[
 (1+2x)^A=\sum_{j=0}^A 2^j\binom Aj x^j.
\]
The largest coefficient of $(1+2x)^2=1+4x+4x^2$ is $4$.
Multiplication by $1+2x$ replaces each coefficient by itself plus
twice the preceding coefficient, increasing the maximum by a factor
of at most $3$. Thus every coefficient of $(1+2x)^A$ is at most
$4\cdot3^{A-2}\le4^{A-1}$.
\end{proof}

\section{Deferred construction of matching moments}
\label{proof:5.4}
Our last missing proof is for a lemma in \Cref{sec:classicalmomentconstruction}.

\momentconstructionlemma*
\begin{proof}[Proof of \Cref{lem:quadratic-support}]
Let $d=r-2$, $B=\lceil2/\eta\rceil$, and
$A=1+\lceil16Br^2/\eta\rceil\le48r^2/\eta^2$. Define
\[
 E_d=\min_{q: \deg q\le d}\max_{a\in[A]}|a^{-1}-q(a)|\le\frac12.
\]
The bound $E_d\le1/2$ follows by choosing the constant
polynomial $q(a)=1/2$. %
A polynomial $q$ with degree at most $d$ that minimizes the above satisfies $|q(a)|\le a^{-1}+E_d\le3/2<2$ for $a\in[A]$.
Applying Lemma~\ref{lem:taylor} to $q(1+z)/2$, with keeping only the constant
Taylor term, and comparing the approximation errors at $1$
and $B$, gives us
\[
\begin{aligned}
 |q(1)-q(B)|&\le\frac{8(B-1)d^2}{A-1}\le\frac\eta2,\\
 2E_d&\ge1-\frac1B-|q(1)-q(B)|\ge1-\eta.
\end{aligned}
\]

Let us write $q$ in the monomial basis, now taking the linear programming
dual of the constraints $|a^{-1}-q(a)|\le t$, we have
\[
\begin{aligned}
 E_d=\max_{u,v\in\mathbb R_{\ge0}^A}\;&
       \sum_{a=1}^A\frac{u_a-v_a}{a}\\
 \text{subject to}\;&\sum_{a=1}^A(u_a+v_a)=1,\\
 &\sum_{a=1}^A(u_a-v_a)a^j=0
       \qquad(0\le j\le d).
\end{aligned}
\]
For an optimal solution, the $j=0$ constraint implies that $\sum_a u_a=\sum_a v_a=1/2$. Hence $\nu_0(a)=2u_a$ and
$\nu_1(a)=2v_a$ are probability distributions with matching moments
through degree $d$. For $\theta_b:=\sum_a\nu_b(a)/a$, the objective gives
$\theta_0-\theta_1=2E_d\ge1-\eta$.
Since $0<\theta_b\le1$, we have $\theta_0\ge1-\eta$ and
$\theta_1\le\eta$. Define
\[
 \Pr[X_b=a]=\frac{\nu_b(a)}{a\theta_b}.
\]
Then $\E[X_b]=1/\theta_b$ and
$\E[X_b^j]/\E[X_b]=\sum_a\nu_b(a)a^{j-1}$, proving the claims.
\end{proof}

\end{document}